\documentclass[prl,twocolumn,nofootinbib,aps,tightenlines,preprintnumbers,notitlepage,longbibliography,superscriptaddress]{revtex4-1}
\usepackage{graphicx}
\usepackage{amsmath}
\usepackage{amsfonts}
\usepackage[colorlinks=true,linktocpage=true,urlcolor=blue,citecolor=blue,linkcolor=blue]{hyperref}

\newcommand{\nhat}{\hat{\boldsymbol{n}}}

\usepackage{amsmath}
\usepackage{amsfonts}

\usepackage{tabularx}
\newcolumntype{C}{>{\centering\arraybackslash}X}
\newcolumntype{R}{>{\raggedleft\arraybackslash}X}

\usepackage{dsfont}

\usepackage[usenames, dvipsnames]{color}
\usepackage[normalem]{ulem}
\usepackage{xcolor}

\newcommand{\dd}{{\rm d}}

\def\nhat{\hat{\boldsymbol{n}}}
\def\rhat{\hat{\mathbf{r}}}

\newcommand{\br}{\boldsymbol{r}}

\newcommand{\be}{\begin{eqnarray}}

\newcommand{\ee}{\end{eqnarray}}
\newcommand{\brp}{\boldsymbol{r}\!_\perp}
\newcommand{\balpha}{\boldsymbol{\alpha}}
\definecolor{colorA}{HTML}{1E90FF}
\definecolor{colorB}{HTML}{228B22}
\definecolor{colorC}{HTML}{FF7F00}
\definecolor{colorD}{HTML}{4B0082}
\definecolor{colorE}{HTML}{B22222}

\definecolor{lgreen}{HTML}{32CD32}
\definecolor{lgray}{HTML}{D3D3D3}
\definecolor{dblue}{HTML}{1E90FF}
\definecolor{dblue}{HTML}{1E90FF}
\definecolor{orange}{HTML}{FF4500}
\definecolor{indigo}{HTML}{4B0082}

\definecolor{teal}{HTML}{008080}
\definecolor{firebrick}{HTML}{B22222}
\definecolor{salmon}{HTML}{FA8072}
\definecolor{darkgreen}{HTML}{006400}

\newcommand{\jhu}{William H. Miller III Department of Physics and Astronomy, Johns Hopkins University, Baltimore, MD 21218, USA}
\newcommand{\CnuB}{{C$\nu$B}}

\graphicspath{ {plots/} }

\makeatletter
\def\doauthor#1#2#3{%
  \ignorespaces#1\unskip
  \begingroup
   #3%
  \@if@empty{#2}{\@listcomma\endgroup{}{}}{\endgroup{\comma@space}{}\frontmatter@footnote{#2}}%
  \space \@listand
}%
\makeatother

\makeatletter
\def\@ssect@ltx#1#2#3#4#5#6[#7]#8{%
  \def\H@svsec{\phantomsection}%
  \@tempskipa #5\relax
  \@ifdim{\@tempskipa>\z@}{%
    \begingroup
      \interlinepenalty \@M
      #6{%
       \@ifundefined{@hangfroms@#1}{\@hang@froms}{\csname @hangfroms@#1\endcsname}%
       {\hskip#3\relax\H@svsec}{#8}%
      }%
      \@@par
    \endgroup
    \@ifundefined{#1smark}{\@gobble}{\csname #1smark\endcsname}{#7}%
  }{%
    \def\@svsechd{%
      #6{%
       \@ifundefined{@runin@tos@#1}{\@runin@tos}{\csname @runin@tos@#1\endcsname}%
       {\hskip#3\relax\H@svsec}{#8}%
      }%
      \@ifundefined{#1smark}{\@gobble}{\csname #1smark\endcsname}{#7}%
      \addcontentsline{toc}{#1}{\protect\numberline{}#8}%
    }%
  }%
  \@xsect{#5}%
}%
\makeatother

\begin{document}

\title{Unveiling Neutrino Halos with CMB Lensing}

\author{Selim~C.~Hotinli$^{\mathds{H},}$}
\affiliation{\jhu}
\author{Nashwan Sabti$^{\mathds{S},}$}
\affiliation{\jhu}
\author{Jaxon North$^{\mathds{N},}$}
\affiliation{\jhu}
\author{Marc Kamionkowski$^{\mathds{K},}$}
\affiliation{\jhu}

\def\thefootnote{$\mathds{H}$\hspace{0.7pt}}\footnotetext{\href{mailto:shotinl1@jhu.edu}{shotinl1@jhu.edu}}
\def\thefootnote{$\mathds{S}$\hspace{1.7pt}}\footnotetext{\href{mailto:nsabti1@jhu.edu}{nsabti1@jhu.edu}}
\def\thefootnote{$\mathds{N}$\hspace{0.5pt}}\footnotetext{\href{mailto:jnorth8@jhu.edu}{jnorth8@jhu.edu}}
\def\thefootnote{$\mathds{K}$}\footnotetext{\href{mailto:kamion@jhu.edu}{kamion@jhu.edu}}
\setcounter{footnote}{0}
\def\thefootnote{\arabic{footnote}}

\begin{abstract}
\noindent The existence of a cosmic neutrino background has been inferred indirectly from cosmological surveys through its effect on the linear-theory evolution of primordial density perturbations, as well as from measurements of the primordial abundances of light elements. Constraints on the masses of the three neutrino species imply that at least two of them move non-relativistically today. As a consequence, non-linear evolution of density perturbations results in the formation of neutrino halos around dark-matter halos. We study whether these neutrino halos can be detected in the foreseeable future through measurements of weak gravitational lensing of the cosmic microwave background, thus providing, possibly, the first beyond-linear-theory signature of cosmic neutrinos.
\end{abstract}

\maketitle
 
\section{Introduction}

A cosmic neutrino background (\CnuB) is an unavoidable consequence of the hot Big-Bang model for the origin of the Universe~\cite{Weinberg:1972kfs, Dodelson:1992km}. Its properties are intricately linked to the physics governing the Universe approximately 1 second after its birth. The detection of the \CnuB\ would therefore mark a substantial leap in unraveling the earliest epoch of our Universe's history.

Direct detection of these neutrinos is notoriously difficult, even by the standards of most of neutrino physics, given that \CnuB\ neutrinos today have energies $\lesssim10^{-3}\,\mathrm{eV}$ and neutrino interaction strengths scale with this energy. As early as 1962, Ref.~\cite{Weinberg:1962zza} proposed an inverse-beta-decay technique to directly detect neutrinos from the \CnuB. Despite past and ongoing efforts in advancing this framework~\cite{Cocco:2007za}, the experimental program is expected to require several more decades to bear fruit~\cite{PTOLEMY:2019hkd, Alvey:2021xmq, PTOLEMY:2022ldz}. While alternative approaches to detect the \CnuB\ have been suggested~\cite{Stodolsky:1974aq, Langacker:1982ih, Duda:2001hd, Eberle:2004ua, Bauer:2021uyj, Yoshimura:2014hfa, Domcke:2017aqj, Alonso:2018dxy, Bauer:2022lri, Arvanitaki:2022oby}, they often face similar challenges in terms of sensitivity. Concurrently, an extensive laboratory program has been established to measure fundamental properties of neutrinos, such as their masses, with the current leading constraints set by the KATRIN experiment, imposing an upper bound on the sum of neutrino masses of $\sum m_\nu < 2.4\,\mathrm{eV}$~\cite{KATRIN:2021uub}.

From a {cosmological point of view}, the inferred properties of neutrinos are {contingent upon} the underlying cosmological model assumed. Within the standard $\Lambda$CDM model, neutrinos are expected to follow a Fermi-Dirac distribution with a temperature of $1.95\,\mathrm{K}$. Within this framework, the effects of the \CnuB\ on the linear-theory evolution of primordial density perturbations have been seen in cosmic microwave background (CMB) observations and galaxy surveys, see e.g.~\cite{Lesgourgues:2006nd} for a review. In addition, measurements of the primordial abundances of light elements have {enabled to probe} the properties of neutrinos in the very early Universe~\cite{Fields:2019pfx}. This has resulted in {measurements of and constraints on the relativistic and non-relativistic neutrino energy densities, respectively,} down to $O(10\%)$ precision~\cite{Planck:2018vyg, Alvey:2021sji}. The latter measurement has {yielded} the {strongest} (but model-dependent) constraint on the sum of neutrino masses to date, setting an upper bound of $\sum m_\nu < 0.12\,\mathrm{eV}$~\cite{Planck:2018vyg}.

The presence of non-zero neutrino masses has interesting implications for the formation of cosmic structures. Measurements of solar and atmospheric neutrinos have shown that at least two of the neutrino mass eigenstates are heavy enough to move non-relativistically today~\cite{Esteban:2020cvm}. As a result, the non-linear evolution of primordial perturbations would lead to the emergence of neutrino halos that are gravitationally bound to dark-matter halos, but exhibit a more diffuse distribution~\cite{Ringwald:2004np,Villaescusa-Navarro:2011loy,Ichiki:2011ue,Zhang:2017ljh, deSalas:2017wtt,Mertsch:2019qjv,Holm:2023rml}. Detecting these neutrino halos through their gravitational effects would constitute a new, albeit indirect, observational signature of the \CnuB. This concept has been explored in the context of cosmic-shear measurements in Ref.~\cite{Villaescusa-Navarro:2011loy}, where it was shown that neutrino halos would introduce only a (sub)percent-level effect on the shear profile, making their detection a technical challenge.

In this paper, we study the potential of detecting neutrino halos around dark-matter halos through their lensing influence on the CMB. On sufficiently small angular scales, the CMB temperature/polarization can be approximated by a gradient across the halos. Lensing of this gradient then induces a dipolar distortion whose detailed pattern is determined by the lens mass distribution, see e.g.~\citep{Seljak:1999zn,Lewis:2005fq,Levy:2023moy,Horowitz:2017iql,Birkinshaw:1983,Hotinli:2018yyc,Hotinli:2020ntd}. We show whether stacking CMB patches around galaxies with future CMB experiments may enable a detection of this subtle effect, and discuss the main challenges that need to be addressed in order to achieve this goal.

Throughout this work, we assume a flat $\Lambda$CDM cosmology and fix the cosmological parameters to the Planck 2018 best-fits~\cite{Planck:2018vyg}: $h = 0.6727$, $\omega_\mathrm{cdm} = 0.1202$, $\omega_\mathrm{b} = 0.02236$, $n_\mathrm{s} = 0.9649$, $\ln\left(10^{10}A_\mathrm{s}\right) = 3.045$, and $\tau_\mathrm{reio} = 0.0544$.

\section{Dark-matter and neutrino halos}

We assume that dark-matter halos of mass $M$ follow a spherically symmetric NFW profile, given by 
\be\label{eq:nfw_prof}
\rho_{\rm DM} = \frac{\rho_\mathrm{s}}{x(1+x)^2}\ ,
\ee
where
\be
\rho_\mathrm{s} = \frac{M}{4\pi r_\mathrm{s}^3\left[\ln(1+c) - c/(1+c)\right]}\ ,
\ee
and $x = cr/r_\mathrm{vir}$. Here, $c$ is the concentration parameter, and $r_\mathrm{vir}$ is the virial radius. We use the concentration-halo relation from Ref.~\cite{Correa:2015dva}, and define the virial radius as $r_\mathrm{vir} = \left(3M/(200\times 4\pi\rho_\mathrm{crit,0})\right)^{1/3}$. Using the Poisson equation, we can then calculate the gravitational potential $\Phi_{\rm DM}= -4\pi G\rho_\mathrm{s} r_\mathrm{s}^2{\ln(1+x)}/{x}$ and its derivative 
\begin{align}
\label{eq:Poisson}
\frac{\partial\Phi_{\rm DM}}{\partial r} &= 4\pi G\rho_\mathrm{s} r_\mathrm{s}\left[\frac{\ln(1+x)}{x^2}-\frac{1}{x(1+x)}\right]\ ,
\end{align}
which we use to compute the deflection of CMB photons caused by the dark matter.

To obtain the contribution from neutrinos to the gravitational potential, we first solve the collisionless Boltzmann equation for the neutrino distribution function. We do this by linearizing the Boltzmann equation, following the approach in Refs.~\cite{Singh:2002de,Ringwald:2004np}, which after solving gives us the neutrino density perturbation. This neutrino density perturbation multiplied by the background density is then interpreted as the neutrino-halo profile. In Fourier-space, the neutrino perturbation $\hat{\delta}_\nu$ reads~\cite{Ringwald:2004np}
\begin{align}
    \label{eq:delta_nu_fourier}
    \hat{\delta}_\nu(\boldsymbol{k},z, M) 
    =\ & 4\pi G\bar{\rho}_{\mathrm{m},0}\!\int_{s_\mathrm{ini}}^{s(z)}\!\!\!\!\mathrm{d}s' a(s')\hat{\delta}_\mathrm{m}(\boldsymbol{k},s',M)(s(z)\!-\!s')  \nonumber\\
    &\qquad\qquad\quad\ \, \times F\left[\frac{k(s(z)\!-\!s')}{m_\nu}\right]\ ,
\end{align}
where $s = \int\mathrm{d}t\, a^{-2}$ (and $a$ the scale factor), $\hat{\delta}_\mathrm{m}$ is the matter density perturbation (which we obtain with \texttt{CLASS}~\cite{Blas:2011rf}), and $F$ is the Fourier transform of the Fermi-Dirac distribution, given by
\begin{align}
    F(x) = \frac{4}{3\zeta(3)}\sum_{n=1}^{\infty}(-1)^{n+1}\frac{n}{(n^2+x^2T_\nu^2)^2}\ ,
\end{align}
with $T_\nu = 1.95(1+z)\,\mathrm{K}$. We define the real-space matter perturbation as $\delta_\mathrm{m} = \rho_\mathrm{DM}/\overline{\rho}_\mathrm{m}$, with $\overline{\rho}_\mathrm{m}$ the matter background density, and use the NFW profile as given above. The neutrino density perturbation in real space can be obtained by Fourier transforming Eq.~\eqref{eq:delta_nu_fourier}
\begin{align}
    \label{eq:deltanu_real}
    \delta_\nu(r, z, M) = \frac{1}{2\pi^2}\int\mathrm{d}k k^2\frac{\sin(kr)}{kr}\hat{\delta}_\nu(k,z,M)\ .
\end{align}
\begin{figure*}[t!]
    \centering    \includegraphics[width=\linewidth]{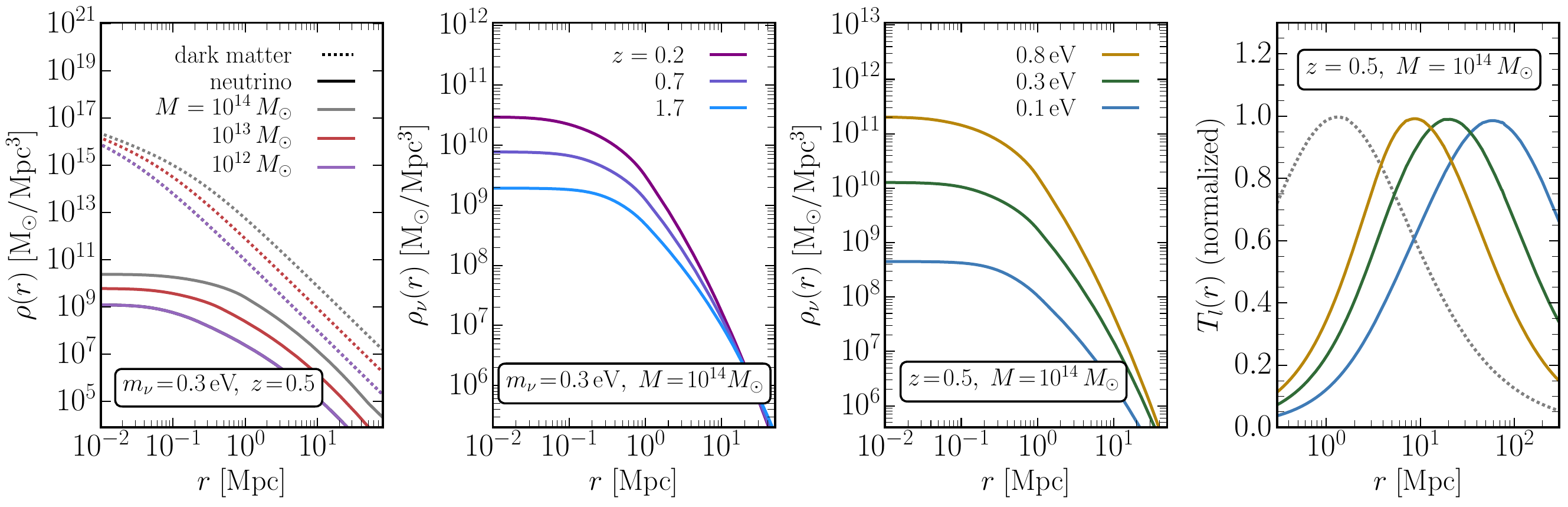}\vspace*{-0.45cm}
    \caption{\textbf{Left three panels:} Density profiles of dark-matter (dotted) and neutrino (solid) halos. Left-most panel shows the density profiles for three dark-matter halo masses at redshift $z=0.5$ and for $m_\nu=0.3\, {\rm eV}$. The second and third panels show the dependence of the neutrino halo profiles on redshift and neutrino mass. \textbf{Rightmost panel:} The normalized lensed CMB temperature profiles from dark matter (dotted) and neutrino (solid) halos. The curves are normalized to unity at their maximum. CMB lensing due to neutrino halos can be seen to peak further away from the halo center compared to the dark-matter induced effect.}\label{fig:FIG1}
\end{figure*}

We note that while strictly speaking this approach is valid only when $\delta_\nu \ll 1$, the linear approximation is generally an underestimation of the true clustering as obtained with simulations~\cite{Singh:2002de,Ringwald:2004np}, which makes our method here more conservative. In what follows, we take Eq.~\eqref{eq:deltanu_real} and define $\rho_\nu(r|z,M) \equiv f_{\nu,\rm nl}\,\overline{\rho}_\nu \delta_\nu(r,z,M)$, where $f_{\nu,\rm nl}$ is a fudge-factor to account for the potential mismatch between our approximation and fits using N-1-body simulations (set to unity throughout), and $\overline{\rho}_\nu$ is the background neutrino density. We then again use the Poisson equation to calculate the neutrino halo potential and its derivative numerically. To assess the impact of the neutrino mass scale on our results, we consider three scenarios\footnote{Note that for $m_\nu^\mathrm{lightest} \gtrsim 0.1\,\mathrm{eV}$ the neutrinos can be treated as degenerate in mass~\cite{Lesgourgues:2012uu}.}: a case with $m_\nu^\mathrm{lightest} = 0.1\,\mathrm{eV}$, a higher mass scenario with $m_\nu^\mathrm{lightest} = 0.3\,\mathrm{eV}$, and the current upper limit set by KATRIN with $m_\nu^\mathrm{lightest} = 0.8\,\mathrm{eV}$~\cite{KATRIN:2021uub}.

We show the matter and neutrino density profiles in the first three panels of Fig.~\ref{fig:FIG1}. The first panel shows the profiles for three different dark-matter halo masses $M=\{10^{12},10^{13},10^{14}\}\, M_\odot$. The neutrino halo density can be seen to be many orders of magnitude smaller -- and smoother near the halo center -- compared to the dark-matter halo.  The second and third panels show the redshift and neutrino-mass dependence of the neutrino-halo profile. The rightmost panel shows normalized CMB lensing profiles due to dark matter (dotted line) and neutrino (solid lines) halos, demonstrating how the lensing signal from neutrinos peaks at larger radii compared to dark matter, and in a way sensitive to neutrino mass.

\section{CMB lensing}

CMB photons from the last-scattering surface get deflected by potential gradients along our line-of-sight direction $\nhat$ (see Ref.~\citep{Lewis:2006fu} for a review). The lensed CMB temperature and polarization satisfy ${X}^l(\nhat)=X^u(\nhat+\balpha(\nhat))$,
where $X\in\{T,Q,U\}$ represents the dimensionless CMB temperature $T$ or the two Stokes parameters $Q$ or $U$ (normalized by the mean CMB temperature). Throughout this work, we will use $l$ ($u$) to denote lensed (unlensed) CMB fluctuations. At leading order in perturbations, the lensed CMB can be expanded as
\begin{equation}\label{eq:T_lens}
\begin{split}
    X^u(\nhat+\balpha(\nhat)) \simeq X^u(\nhat) + \balpha(\nhat)\cdot\boldsymbol{\nabla} X^u(\nhat)\ ,\\
\end{split}
\end{equation}
and the lensing deflection $\balpha$ can be assumed to be a pure gradient $\boldsymbol{\alpha}(\nhat) = \boldsymbol{\nabla}\phi(\nhat)$, such that
\be
    \balpha(\nhat) = -2\, \boldsymbol{\nabla}\!\!\int_0^{\chi_\star} \mathop{\dd \chi} \frac{D_{\rm ds}}{D_{\rm s}} \Phi(\chi\nhat) \ .
\ee
Here, $\Phi(\chi\nhat)$ is the gravitational potential, $\chi$ is the comoving distance, $\chi_\star$ is the comoving distance to the recombination surface, $D_{\rm ds}\equiv\chi_\star-\chi$, and $D_{\rm s}\equiv\chi_\star$. We approximate the gravitational potential near a dark-matter halo to be spherically symmetric around the halo center, $\boldsymbol{\nabla}\Phi(r)=\rhat\,\partial\Phi(r)/\partial r$, and define the unit vector $\hat{\boldsymbol{r}}$ with origin at the halo center. 

The CMB modulations due to lensing sourced by a single halo can be written in the flat-sky approximation as
\be
\begin{split}
X^l(\brp)\!=\!-\frac{4}{c^2}[\boldsymbol{\nabla}X^u(\brp) \!\cdot\!\boldsymbol{r}_\perp]\frac{D_{\rm ds}}{D_{\rm s}}\!\int_{r_\perp}^{\infty}\!\!\!\!\frac{\dd r}{\sqrt{{r}^2-{r}^2_\perp}}\frac{\partial\Phi}{\partial r}\ ,
\end{split}
\ee
where $\brp$ is the transverse radial distance away from the halo center, and $\boldsymbol{\nabla}X^u(\brp)$ is the unlensed CMB fluctuation gradient. Here we assume both the CMB temperature gradient and the parameter $D_{\rm ds}$ are constant in the integration region. Before describing our stacking analysis, we first detail the galaxy survey configuration we consider.

\section{Galaxy survey}

For the galaxy survey, we consider the Vera Rubin Observatory (LSST) deep sample following Ref.~\citep{Ferraro:2022twg}. We assume that the galaxy catalog can be separated into halo mass and redshift bins, and calculate the lensing signal on an evenly-spaced grid of 10 halo mass bins within $\log_{10}M\in[12,15]\, M_\odot$ by 5 redshift bins within $z\in[0.1,3]$. The total galaxy number density at each redshift roughly satisfies  $n(z)=n_{\rm gal}({z}/{z_0})^{2}\exp(-z/z_0)/(2z_0)$, 
where $z_0=0.3$ and we set the galaxy number to roughly twice the number anticipated for the LSST gold sample: $n_{\rm gal}\simeq80\,{\rm arcmin}^{-2}$~\citep{LSSTScience:2009jmu,Ferraro:2022twg}. We compute the anticipated number count of halos in each mass-redshift bin using the standard Sheth–Tormen halo mass function~\citep{Sheth:1999mn}, normalized to the expected total galaxy number $n(z)$ at each redshift in a volume corresponding to a sky coverage appropriate for the LSST survey.

The individual halo masses are expected to be measured imperfectly, with around $40\%$ error in $\ln M$, from combinations of lensing and Sunyaev-Zeldovich (SZ) measurements. In addition, redshift determinations are subject to photo-$z$ errors~\citep{Palmese:2019lkh,Murata:2017zdo,Ballardini:2019wxj}; for example, the anticipated LSST photo-$z$ error is $\sigma_z(z)=0.03(1+z)$~\citep{LSSTScience:2009jmu}. As we show next, using information on halo mass and redshift would lead to a more optimal estimation of the lensing profile, and can potentially break degeneracies between parameters that affect CMB lensing in similar ways. Here, we assume that the effect of the measurement uncertainties of halo mass and redshift can be accounted for by taking galaxy catalog mass-redshift bin sizes to be sufficiently larger than these errors\footnote{Future experiments such as MegaMapper~\citep{Schlegel:2019eqc}, which is expected to have precise redshift information for all halos observable with LSST, could allow for a much finer binning in redshift, significantly increasing the prospects of probing neutrino halos at higher precision.}.

\begin{figure*}[t!]
    \centering
    \includegraphics[width = \textwidth]{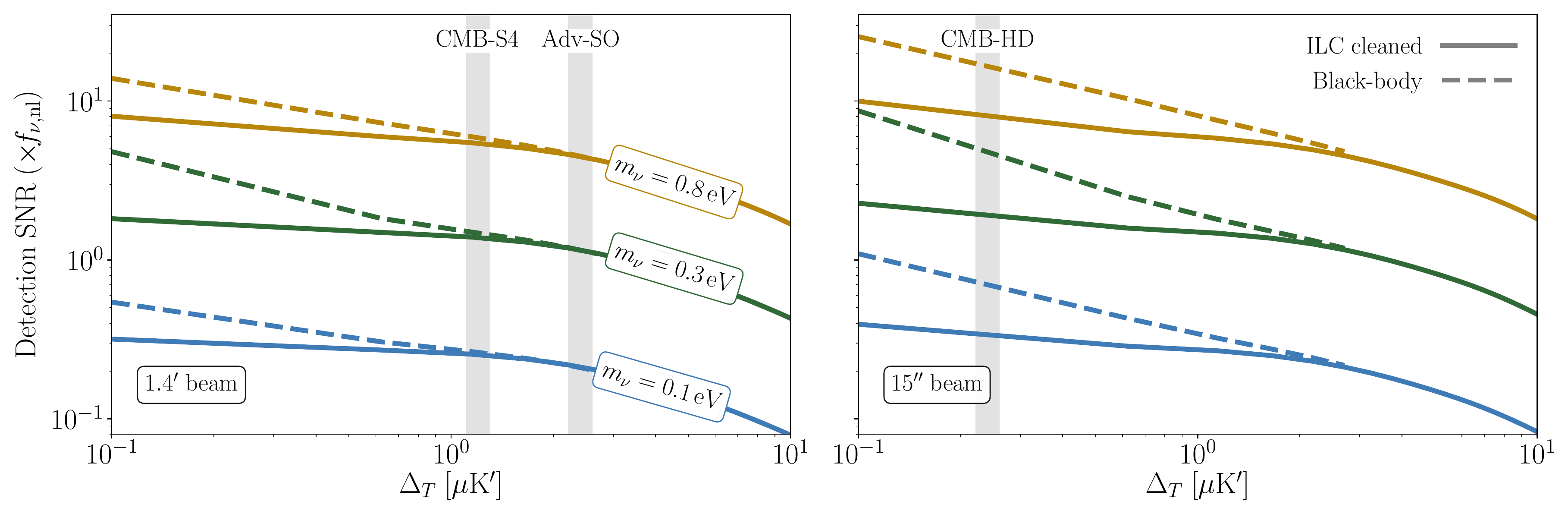}
    \caption{Detection signal-to-noise ratio of neutrino halos for next-generation CMB experiments and a galaxy catalog representative of the LSST deep sample.  
    The dashed lines are obtained by assuming frequency-dependent foregrounds can be removed from maps, leaving the lensed black-body CMB signal and black-body foregrounds only, while the solid ones correspond to the detection SNR assuming standard ILC-cleaning as described in the text. In the left (right) panel, we assume the beam of the residual CMB noise is 1.4-arcminute (15-arcsecond), appropriate to upcoming (future) CMB experiments. The horizontal axis corresponds to the ILC-cleaned white-noise RMS of CMB experiments, with gray shaded regions depicting anticipated noise levels matching CMB-S4/Advanced SO (left panel) and CMB-HD (right panel). The blue, green, and yellow lines in both panels correspond to neutrino masses $m_\nu=\{0.1,0.3,0.8\}\,{\rm eV}$, respectively. Here we set $f_{\rm sky}=0.4$ and $f_{\nu,\rm nl}=1$.}
    \label{fig:FIG2} 
\end{figure*}

\section{Stacking analysis}

To extract the lensing signal, we stack patches of the CMB around dark-matter halos oriented along the unlensed CMB gradients. We apply a matched filter to each patch that returns the unbiased minimum-variance lensing signal to maximize the prospects of detection.

We begin by writing the observed real-space intensity map around a dark-matter halo as composed of the lensing signal, $\boldsymbol{\alpha}(\br)\equiv\hat{\br}_\perp\alpha(\br)$, and all other contributions
\begin{equation}
\tilde{X}^{l}({\boldsymbol{r}})=-\boldsymbol{\nabla}X^u({\boldsymbol{r}})\cdot\hat{\boldsymbol{r}}\!_\perp\,{\alpha}(\boldsymbol{r})+\tilde{X}({\boldsymbol{r}})\ ,
\end{equation} 
where $\alpha(\br)=\alpha_\nu(\br)+\alpha_{\rm DM}(\br)+\alpha_{\rm b}(\br)$ is the total lensing deflection profile including neutrinos, dark matter, and baryons, respectively. We assume that the observed lensed CMB maps can be filtered to get unbiased and minimum-variance estimates for the norm of the unlensed CMB gradient  $\widehat{|\boldsymbol{\nabla} X^u|}$ for each halo. The corresponding estimator noise $N^X\equiv\langle(\widehat{|\boldsymbol{\nabla} X^u|}-{|\boldsymbol{\nabla} X^u|})^2\rangle$ from neutrino-lensing can be shown to satisfy~\citep{Hotinli:2020ntd}
\be
N^X_\nu(M,z)=\frac{1}{2}\int\frac{\dd^2\boldsymbol{\ell}}{(2\pi)^2} \frac{|{\alpha}_\nu(\boldsymbol{\ell})|^2}{\tilde{C}_\ell^{XX}}\ ,
\ee
where $\alpha_\nu(\boldsymbol{\ell})$ is the neutrino-halo lensing deflection profile in Fourier space, and $\tilde{C}_\ell^{XX}$ is the ILC-cleaned CMB spectrum, consisting of the lensed primary CMB signal, foregrounds, and CMB experiment noise\footnote{Here we assumed that the neutrino-halo lensing can be isolated from the CMB patches. In practice, lensing due to dark-matter halos and baryons will bias the measurement of the unlensed CMB gradients, roughly proportional to ${\sim}|\alpha_{\nu}(\boldsymbol{\ell})\alpha_{\rm DM}(\boldsymbol{\ell})|$ and ${\sim}|\alpha_{\nu}(\boldsymbol{\ell})\alpha_{\rm b}(\boldsymbol{\ell})|$, affecting prospects to unambiguously detect the neutrino-halo lensing. We take a fist step towards investigating this bias in what follows.}.

We calculate the lensed CMB black-body temperature and polarization power-spectra using \texttt{CAMB}~\cite{CAMB}. The Stokes parameters spectra $C_\ell^{QQ}$ and $C_{\ell}^{UU}$ are derived from $C_\ell^{EE}$ and $C_\ell^{BB}$. We account for the foregrounds following the prescription in the appendix of Ref.~\citep{Hotinli:2021hih}. In particular, we add the following contributions to the temperature power spectrum: the frequency-dependent cosmic infrared background, the thermal Sunyaev-Zeldovich (tSZ) effect, radio sources, and the reionization and late-time kinetic Sunyaev-Zeldovich (kSZ) effects (which are black-body). We assume that the polarization spectra are foreground-free. The CMB noise is modeled at each frequency channel and includes the pink and white noise components as $N_\ell=\Delta_X^2\exp\left[\ell(\ell+1){\theta_{\rm FWHM}^2}/{(8\log(2))}\right][1+(\ell_{\rm knee}/\ell)^{\alpha_{\rm knee}}]$, with $\theta_{\rm FWHM}$ the beam full-width at half-maximum (FWHM), $\Delta_X$ the noise root mean square (RMS), and $\{\alpha_{\rm knee},\ell_{\rm knee}\} = \{-3, 100\}$ parameters that model the effect of the Earth's atmosphere for {upcoming ground-based experiments}. For our forecast, we consider next-generation CMB experiments, ranging from the upcoming Simons Observatory (SO)~\cite{SimonsObservatory:2019qwx} to the future CMB-S4~\cite{CMB-S4:2016ple,Abazajian:2019eic}, Advanced SO~\cite{SimonsObservatory:2019qwx} and CMB-HD~\cite{Sehgal:2019ewc,CMB-HD:2022bsz}, whose survey parameters we take from table~2 of Ref.~\citep{Hotinli:2021hih}. We combine the anticipated CMB observations at different frequency channels to calculate the minimum-variance standard internal linear combination\footnote{{Ongoing work on mitigating the effects of foregrounds for lensing measurements, as well as delensing, and development of more optimal lensing-potential reconstruction techniques suggests the fidelity of lensing reconstruction in the future may improve significantly compared to the ILC-cleaned CMB spectrum $\tilde{C}_\ell^{XX}$ which we consider here~(see e.g.~\citep{Namikawa:2012pe, Millea:2017fyd, Horowitz:2017iql, Hadzhiyska:2019cle, Darwish:2021ycf, Sailer:2020lal, Sailer:2022jwt, 2022arXiv221108990P, Foreman:2022ves, 2022arXiv220507368L, Han:2021vtm, Hotinli:2021umk}).}} (ILC) cleaned~\citep{Tegmark:2003ve} CMB spectrum $\tilde{C}_\ell^{XX}$.

The estimated signal-to-noise ratio (SNR) from CMB field $X$ for a halo of mass $M$ at redshift $z$ is then
$\mathrm{SNR} = \sqrt{{\langle|\boldsymbol{\nabla} X^u|^2\rangle}/{N^X_\nu(M,z)}}$, where we set the fiducial values of the dimensionless CMB gradient RMS $\{\sqrt{\langle|\boldsymbol{\nabla}T^u |^2\rangle},\sqrt{\langle| \boldsymbol{\nabla} Q^u|^2 \rangle},\sqrt{\langle|\boldsymbol{\nabla}U^u |^2\rangle}\}$ equal to\footnote{Or, equivalently, $\{\sqrt{\langle|\boldsymbol{\nabla}T^u |^2\rangle},\sqrt{\langle| \boldsymbol{\nabla} Q^u|^2 \rangle},\sqrt{\langle|\boldsymbol{\nabla}U^u |^2\rangle}\}$ = $\{13, 1.3, 1.3\}$ in ${\mu {\rm K}\,{\rm arcmin}^{-1}}$ units.} $\{4.7,0.47,0.47\}\!\times\!10^{-6}$.
We first calculate the single-halo detection SNRs from the three components $X\in\{T,Q,U\}$ for each halo mass and redshift bin in our grid. We then weight the SNR values with the number count of halos at each mass-redshift bin expected from LSST. Finally, we sum these in inverse quadrature to obtain the total detection SNR from the catalog.

\section{Results}

The SNR is shown in Fig.~\ref{fig:FIG2} as a function of the CMB temperature white noise amplitude after ILC-cleaning. We also consider forecasts assuming all frequency-dependent foregrounds can be removed from the CMB (dashed lines), leaving only the black-body CMB signal and foregrounds. We find that at the present cosmological mass limit within $\Lambda$CDM, $\sum m_\nu <0.12\,\mathrm{eV}$, the SNR is too low for the neutrino-halo-lensing effect to be detectable in the near future. We emphasize, however, that although neutrino masses are very small in the standard $\Lambda$CDM model, it is reasonable to question, given residual tensions like those with $H_0$ and $S_8$~\cite{Verde:2019ivm,DiValentino:2020vvd, Perivolaropoulos:2021jda}, whether there may be some new physics missing. It is also reasonable to surmise that such new physics might allow higher neutrino masses or colder neutrinos. Thus, although the prospects for detection of these neutrino halos is not promising with the nominal current bound and our conservative assumptions for the halo profiles, the search may still be warranted given persistent issues with $\Lambda$CDM and the likelihood that full calculations will indicate denser neutrino halos than we have assumed here (for example, the fudge factor could be as high as a factor of ${\sim}5$ within $\Lambda$CDM~\cite{Ringwald:2004np} or even higher beyond~\cite{Alvey:2021xmq}). Nevertheless, with our current setup, we see that larger neutrino masses lead to a significant increase in the SNR, suggesting measurements of CMB lensing could potentially be used for putting upper bounds on neutrino masses. Notably, future cosmological surveys may have the sensitivity to place such an upper bound to high significance if the frequency-dependent foregrounds can be cleaned efficiently from the CMB.

\section{Discussion and Conclusion}

In this work, we considered a new way to detect cosmic neutrinos beyond their linear-theory effects. In particular, we studied the statistical power of upcoming CMB and large-scale structure surveys to detect neutrino halos from oriented stacking of lensed CMB temperature and polarization patches around galaxy clusters. We showed that {this could be feasible} if the neutrino masses are sufficiently large.

So far, we have omitted uncertainties in the modeling of the dark-matter halo profiles and the effects of baryonic matter on the lensing profiles, both of which would influence the SNR of an unambiguous neutrino-halo detection. While the lensing signal caused by neutrinos have a different shape compared to dark matter (see rightmost panel in Fig.~\ref{fig:FIG1}) and baryons (see e.g.~\citep{Peirani:2016qvp,Sorini:2021eac, Chung:2019bsk}), we anticipate that joint modeling of these effects may make it somewhat more difficult to isolate the impact of neutrinos. As a preliminary analysis, we assess the reduction in the detection {SNR of neutrino halos} when the dark-matter halo mass $M$ and concentration parameter $c$ are modeled as free parameters. We perform an ensemble-statistics analysis where we calculate the information matrix $\mathcal{F}_{\alpha\beta}^{(z)} = \sum_{\ell}\!\frac{f_{\rm sky}({2\ell+1})}{2}[\partial_\alpha C_{\ell}^{TT,\nu(z)}\mathcal{C}^{-1}\partial_\beta C_{\ell}^{TT,\nu(z)}]$. Here, $\tilde{C}_\ell^{TT,\nu(z)}$
is the neutrino-induced lensing profile at redshift $z$, $\mathcal{C}=\tilde{C}_\ell^{TT,\nu(z)}+N_\nu^{T}$ is the corresponding lensing estimator covariance, and indices $\alpha,\beta$ for partial derivatives vary over parameters $\{m_\nu,M,c\}$. For a given halo of mass $M\sim10^{13}\,M_\odot$ and $m_\nu=0.2\,{\rm eV}$, we calculate $\mathcal{F}_{\alpha\beta}^{(z)}$ at each of the five redshift bins we consider within $z\in[0.1,3.0]$ . We then sum these matrices after inverse-weighting them by the number-count of halos anticipated to be in the LSST catalog in these mass-redshift bins. We find that marginalizing over dark-matter halo parameters leads to a modest reduction in the neutrino-halo detection SNR by ${\sim}20\%$ compared to parameterizing the lensing signal with $m_\nu$ alone, i.e., $\sqrt{(\mathcal{F}^{-1})_{m_\nu m_\nu}/(\mathcal{F}_{m_\nu m_\nu})^{-1}}\sim0.8$. In a follow-up work, we will investigate the impact of modeling dark-matter and baryonic effects jointly with neutrinos in further detail. 

Another assumption we made was that the unlensed CMB temperature and polarization gradients can be known perfectly when aligning the patches. This is a reasonable assumption as the unlensed CMB gradients are coherent on small scales due to Silk damping (over ${\sim}90$ percent of the RMS of $\langle|\boldsymbol{\nabla}X^u|^2\rangle$ comes from scales $\ell\lesssim1000$~\citep{Hui:2007cu}) and the observed lensed CMB fields can be delensed very effectively for upcoming CMB surveys that cover large scales (see e.g.~\citep{Hotinli:2021umk,Green:2016cjr}). Nevertheless, in practice, our knowledge of the unlensed CMB gradients will not be perfect due to residual lensing and other interactions of CMB photons with the intervening structure along their trajectories. We leave a more detailed analysis of this effect on the neutrino halo detection prospects to upcoming work. 

Finally, in computing the neutrino halo profile, we solved the linearized Boltzmann equation, which underestimates the true clustering magnitude of massive neutrinos. To {enhance the accuracy} of our results, a more advanced method, such as one based on N-1-body simulations, would capture the non-linear behavior and result in a clustering boost by a factor of a few. We will implement such an improved {methodology} in future work to allow us to refine our findings and obtain more precise results.

The relic neutrino background is an integral part of our cosmological model, yet its direct detection has remained elusive. While laboratory experiments are currently being designed to directly detect these species, cosmology has already provided us with probes to indirectly infer their existence and properties. Neutrinos, being massive, cluster together to create neutrino halos, a mechanism that extends beyond their linear-theory effect in cosmology. In this study, we investigated the prospects of detecting neutrino halos through a joint analysis of CMB lensing and galaxy surveys. Such a detection {would open up new windows} to the origin of our Universe.\\

\noindent \textit{\textbf{Acknowledgements}.\ ---} We thank Niayesh Afshordi, Neal Dalal, Miguel Escudero, Matthew Johnson, Neelima Sehgal, and Kendrick Smith for useful discussions. SCH and NS were supported by a Horizon Fellowship from Johns Hopkins University. MK was supported by NSF Grant No.\ 2112699 and the Simons Foundation.

\bibliography{biblio}

\begin{thebibliography}{76}%
\makeatletter
\providecommand \@ifxundefined [1]{%
 \@ifx{#1\undefined}
}%
\providecommand \@ifnum [1]{%
 \ifnum #1\expandafter \@firstoftwo
 \else \expandafter \@secondoftwo
 \fi
}%
\providecommand \@ifx [1]{%
 \ifx #1\expandafter \@firstoftwo
 \else \expandafter \@secondoftwo
 \fi
}%
\providecommand \natexlab [1]{#1}%
\providecommand \enquote  [1]{``#1''}%
\providecommand \bibnamefont  [1]{#1}%
\providecommand \bibfnamefont [1]{#1}%
\providecommand \citenamefont [1]{#1}%
\providecommand \href@noop [0]{\@secondoftwo}%
\providecommand \href [0]{\begingroup \@sanitize@url \@href}%
\providecommand \@href[1]{\@@startlink{#1}\@@href}%
\providecommand \@@href[1]{\endgroup#1\@@endlink}%
\providecommand \@sanitize@url [0]{\catcode `\\12\catcode `\$12\catcode
  `\&12\catcode `\#12\catcode `\^12\catcode `\_12\catcode `\%12\relax}%
\providecommand \@@startlink[1]{}%
\providecommand \@@endlink[0]{}%
\providecommand \url  [0]{\begingroup\@sanitize@url \@url }%
\providecommand \@url [1]{\endgroup\@href {#1}{\urlprefix }}%
\providecommand \urlprefix  [0]{URL }%
\providecommand \Eprint [0]{\href }%
\providecommand \doibase [0]{http://dx.doi.org/}%
\providecommand \selectlanguage [0]{\@gobble}%
\providecommand \bibinfo  [0]{\@secondoftwo}%
\providecommand \bibfield  [0]{\@secondoftwo}%
\providecommand \translation [1]{[#1]}%
\providecommand \BibitemOpen [0]{}%
\providecommand \bibitemStop [0]{}%
\providecommand \bibitemNoStop [0]{.\EOS\space}%
\providecommand \EOS [0]{\spacefactor3000\relax}%
\providecommand \BibitemShut  [1]{\csname bibitem#1\endcsname}%
\let\auto@bib@innerbib\@empty
\bibitem [{\citenamefont {Weinberg}(1972)}]{Weinberg:1972kfs}%
  \BibitemOpen
  \bibfield  {author} {\bibinfo {author} {\bibfnamefont {Steven}\ \bibnamefont
  {Weinberg}},\ }\href@noop {} {\emph {\bibinfo {title} {{Gravitation and
  Cosmology}: {Principles and Applications of the General Theory of
  Relativity}}}}\ (\bibinfo  {publisher} {John Wiley and Sons},\ \bibinfo
  {address} {New York},\ \bibinfo {year} {1972})\BibitemShut {NoStop}%
\bibitem [{\citenamefont {Dodelson}\ and\ \citenamefont
  {Turner}(1992)}]{Dodelson:1992km}%
  \BibitemOpen
  \bibfield  {author} {\bibinfo {author} {\bibfnamefont {Scott}\ \bibnamefont
  {Dodelson}}\ and\ \bibinfo {author} {\bibfnamefont {Michael~S.}\ \bibnamefont
  {Turner}},\ }\bibfield  {title} {\enquote {\bibinfo {title} {{Nonequilibrium
  neutrino statistical mechanics in the expanding universe}},}\ }\href
  {\doibase 10.1103/PhysRevD.46.3372} {\bibfield  {journal} {\bibinfo
  {journal} {Phys. Rev. D}\ }\textbf {\bibinfo {volume} {46}},\ \bibinfo
  {pages} {3372--3387} (\bibinfo {year} {1992})}\BibitemShut {NoStop}%
\bibitem [{\citenamefont {Weinberg}(1962)}]{Weinberg:1962zza}%
  \BibitemOpen
  \bibfield  {author} {\bibinfo {author} {\bibfnamefont {Steven}\ \bibnamefont
  {Weinberg}},\ }\bibfield  {title} {\enquote {\bibinfo {title} {{Universal
  Neutrino Degeneracy}},}\ }\href {\doibase 10.1103/PhysRev.128.1457}
  {\bibfield  {journal} {\bibinfo  {journal} {Phys. Rev.}\ }\textbf {\bibinfo
  {volume} {128}},\ \bibinfo {pages} {1457--1473} (\bibinfo {year}
  {1962})}\BibitemShut {NoStop}%
\bibitem [{\citenamefont {Cocco}\ \emph {et~al.}(2007)\citenamefont {Cocco},
  \citenamefont {Mangano},\ and\ \citenamefont {Messina}}]{Cocco:2007za}%
  \BibitemOpen
  \bibfield  {author} {\bibinfo {author} {\bibfnamefont {Alfredo~G.}\
  \bibnamefont {Cocco}}, \bibinfo {author} {\bibfnamefont {Gianpiero}\
  \bibnamefont {Mangano}}, \ and\ \bibinfo {author} {\bibfnamefont {Marcello}\
  \bibnamefont {Messina}},\ }\bibfield  {title} {\enquote {\bibinfo {title}
  {{Probing low energy neutrino backgrounds with neutrino capture on beta
  decaying nuclei}},}\ }\href {\doibase 10.1088/1475-7516/2007/06/015}
  {\bibfield  {journal} {\bibinfo  {journal} {JCAP}\ }\textbf {\bibinfo
  {volume} {06}},\ \bibinfo {pages} {015} (\bibinfo {year} {2007})},\ \Eprint
  {http://arxiv.org/abs/hep-ph/0703075} {arXiv:hep-ph/0703075} \BibitemShut
  {NoStop}%
\bibitem [{\citenamefont {Betti}\ \emph {et~al.}(2019)\citenamefont {Betti}
  \emph {et~al.}}]{PTOLEMY:2019hkd}%
  \BibitemOpen
  \bibfield  {author} {\bibinfo {author} {\bibfnamefont {M.~G.}\ \bibnamefont
  {Betti}} \emph {et~al.} (\bibinfo {collaboration} {PTOLEMY}),\ }\bibfield
  {title} {\enquote {\bibinfo {title} {{Neutrino physics with the PTOLEMY
  project: active neutrino properties and the light sterile case}},}\ }\href
  {\doibase 10.1088/1475-7516/2019/07/047} {\bibfield  {journal} {\bibinfo
  {journal} {JCAP}\ }\textbf {\bibinfo {volume} {07}},\ \bibinfo {pages} {047}
  (\bibinfo {year} {2019})},\ \Eprint {http://arxiv.org/abs/1902.05508}
  {arXiv:1902.05508 [astro-ph.CO]} \BibitemShut {NoStop}%
\bibitem [{\citenamefont {Alvey}\ \emph
  {et~al.}(2022{\natexlab{a}})\citenamefont {Alvey}, \citenamefont {Escudero},
  \citenamefont {Sabti},\ and\ \citenamefont {Schwetz}}]{Alvey:2021xmq}%
  \BibitemOpen
  \bibfield  {author} {\bibinfo {author} {\bibfnamefont {James}\ \bibnamefont
  {Alvey}}, \bibinfo {author} {\bibfnamefont {Miguel}\ \bibnamefont
  {Escudero}}, \bibinfo {author} {\bibfnamefont {Nashwan}\ \bibnamefont
  {Sabti}}, \ and\ \bibinfo {author} {\bibfnamefont {Thomas}\ \bibnamefont
  {Schwetz}},\ }\bibfield  {title} {\enquote {\bibinfo {title} {{Cosmic
  neutrino background detection in large-neutrino-mass cosmologies}},}\ }\href
  {\doibase 10.1103/PhysRevD.105.063501} {\bibfield  {journal} {\bibinfo
  {journal} {Phys. Rev. D}\ }\textbf {\bibinfo {volume} {105}},\ \bibinfo
  {pages} {063501} (\bibinfo {year} {2022}{\natexlab{a}})},\ \Eprint
  {http://arxiv.org/abs/2111.14870} {arXiv:2111.14870 [hep-ph]} \BibitemShut
  {NoStop}%
\bibitem [{\citenamefont {Apponi}\ \emph {et~al.}(2022)\citenamefont {Apponi}
  \emph {et~al.}}]{PTOLEMY:2022ldz}%
  \BibitemOpen
  \bibfield  {author} {\bibinfo {author} {\bibfnamefont {A.}~\bibnamefont
  {Apponi}} \emph {et~al.} (\bibinfo {collaboration} {PTOLEMY}),\ }\bibfield
  {title} {\enquote {\bibinfo {title} {{Heisenberg\textquoteright{}s
  uncertainty principle in the PTOLEMY project: A theory update}},}\ }\href
  {\doibase 10.1103/PhysRevD.106.053002} {\bibfield  {journal} {\bibinfo
  {journal} {Phys. Rev. D}\ }\textbf {\bibinfo {volume} {106}},\ \bibinfo
  {pages} {053002} (\bibinfo {year} {2022})},\ \Eprint
  {http://arxiv.org/abs/2203.11228} {arXiv:2203.11228 [hep-ph]} \BibitemShut
  {NoStop}%
\bibitem [{\citenamefont {Stodolsky}(1975)}]{Stodolsky:1974aq}%
  \BibitemOpen
  \bibfield  {author} {\bibinfo {author} {\bibfnamefont {Leo}\ \bibnamefont
  {Stodolsky}},\ }\bibfield  {title} {\enquote {\bibinfo {title} {{Speculations
  on Detection of the Neutrino Sea}},}\ }\href {\doibase
  10.1103/PhysRevLett.34.110} {\bibfield  {journal} {\bibinfo  {journal} {Phys.
  Rev. Lett.}\ }\textbf {\bibinfo {volume} {34}},\ \bibinfo {pages} {110}
  (\bibinfo {year} {1975})},\ \bibinfo {note} {[Erratum: Phys.Rev.Lett. 34, 508
  (1975)]}\BibitemShut {NoStop}%
\bibitem [{\citenamefont {Langacker}\ \emph {et~al.}(1983)\citenamefont
  {Langacker}, \citenamefont {Leveille},\ and\ \citenamefont
  {Sheiman}}]{Langacker:1982ih}%
  \BibitemOpen
  \bibfield  {author} {\bibinfo {author} {\bibfnamefont {Paul}\ \bibnamefont
  {Langacker}}, \bibinfo {author} {\bibfnamefont {Jacques~P.}\ \bibnamefont
  {Leveille}}, \ and\ \bibinfo {author} {\bibfnamefont {Jon}\ \bibnamefont
  {Sheiman}},\ }\bibfield  {title} {\enquote {\bibinfo {title} {{On the
  Detection of Cosmological Neutrinos by Coherent Scattering}},}\ }\href
  {\doibase 10.1103/PhysRevD.27.1228} {\bibfield  {journal} {\bibinfo
  {journal} {Phys. Rev. D}\ }\textbf {\bibinfo {volume} {27}},\ \bibinfo
  {pages} {1228} (\bibinfo {year} {1983})}\BibitemShut {NoStop}%
\bibitem [{\citenamefont {Duda}\ \emph {et~al.}(2001)\citenamefont {Duda},
  \citenamefont {Gelmini},\ and\ \citenamefont {Nussinov}}]{Duda:2001hd}%
  \BibitemOpen
  \bibfield  {author} {\bibinfo {author} {\bibfnamefont {Gintaras}\
  \bibnamefont {Duda}}, \bibinfo {author} {\bibfnamefont {Graciela}\
  \bibnamefont {Gelmini}}, \ and\ \bibinfo {author} {\bibfnamefont {Shmuel}\
  \bibnamefont {Nussinov}},\ }\bibfield  {title} {\enquote {\bibinfo {title}
  {{Expected signals in relic neutrino detectors}},}\ }\href {\doibase
  10.1103/PhysRevD.64.122001} {\bibfield  {journal} {\bibinfo  {journal} {Phys.
  Rev. D}\ }\textbf {\bibinfo {volume} {64}},\ \bibinfo {pages} {122001}
  (\bibinfo {year} {2001})},\ \Eprint {http://arxiv.org/abs/hep-ph/0107027}
  {arXiv:hep-ph/0107027} \BibitemShut {NoStop}%
\bibitem [{\citenamefont {Eberle}\ \emph {et~al.}(2004)\citenamefont {Eberle},
  \citenamefont {Ringwald}, \citenamefont {Song},\ and\ \citenamefont
  {Weiler}}]{Eberle:2004ua}%
  \BibitemOpen
  \bibfield  {author} {\bibinfo {author} {\bibfnamefont {Birgit}\ \bibnamefont
  {Eberle}}, \bibinfo {author} {\bibfnamefont {Andreas}\ \bibnamefont
  {Ringwald}}, \bibinfo {author} {\bibfnamefont {Liguo}\ \bibnamefont {Song}},
  \ and\ \bibinfo {author} {\bibfnamefont {Thomas~J.}\ \bibnamefont {Weiler}},\
  }\bibfield  {title} {\enquote {\bibinfo {title} {{Relic neutrino absorption
  spectroscopy}},}\ }\href {\doibase 10.1103/PhysRevD.70.023007} {\bibfield
  {journal} {\bibinfo  {journal} {Phys. Rev. D}\ }\textbf {\bibinfo {volume}
  {70}},\ \bibinfo {pages} {023007} (\bibinfo {year} {2004})},\ \Eprint
  {http://arxiv.org/abs/hep-ph/0401203} {arXiv:hep-ph/0401203} \BibitemShut
  {NoStop}%
\bibitem [{\citenamefont {Bauer}\ and\ \citenamefont
  {Shergold}(2021)}]{Bauer:2021uyj}%
  \BibitemOpen
  \bibfield  {author} {\bibinfo {author} {\bibfnamefont {Martin}\ \bibnamefont
  {Bauer}}\ and\ \bibinfo {author} {\bibfnamefont {Jack~D.}\ \bibnamefont
  {Shergold}},\ }\bibfield  {title} {\enquote {\bibinfo {title} {{Relic
  neutrinos at accelerator experiments}},}\ }\href {\doibase
  10.1103/PhysRevD.104.083039} {\bibfield  {journal} {\bibinfo  {journal}
  {Phys. Rev. D}\ }\textbf {\bibinfo {volume} {104}},\ \bibinfo {pages}
  {083039} (\bibinfo {year} {2021})},\ \Eprint
  {http://arxiv.org/abs/2104.12784} {arXiv:2104.12784 [hep-ph]} \BibitemShut
  {NoStop}%
\bibitem [{\citenamefont {Yoshimura}\ \emph {et~al.}(2015)\citenamefont
  {Yoshimura}, \citenamefont {Sasao},\ and\ \citenamefont
  {Tanaka}}]{Yoshimura:2014hfa}%
  \BibitemOpen
  \bibfield  {author} {\bibinfo {author} {\bibfnamefont {M.}~\bibnamefont
  {Yoshimura}}, \bibinfo {author} {\bibfnamefont {N.}~\bibnamefont {Sasao}}, \
  and\ \bibinfo {author} {\bibfnamefont {M.}~\bibnamefont {Tanaka}},\
  }\bibfield  {title} {\enquote {\bibinfo {title} {{Experimental method of
  detecting relic neutrino by atomic de-excitation}},}\ }\href {\doibase
  10.1103/PhysRevD.91.063516} {\bibfield  {journal} {\bibinfo  {journal} {Phys.
  Rev. D}\ }\textbf {\bibinfo {volume} {91}},\ \bibinfo {pages} {063516}
  (\bibinfo {year} {2015})},\ \Eprint {http://arxiv.org/abs/1409.3648}
  {arXiv:1409.3648 [hep-ph]} \BibitemShut {NoStop}%
\bibitem [{\citenamefont {Domcke}\ and\ \citenamefont
  {Spinrath}(2017)}]{Domcke:2017aqj}%
  \BibitemOpen
  \bibfield  {author} {\bibinfo {author} {\bibfnamefont {Valerie}\ \bibnamefont
  {Domcke}}\ and\ \bibinfo {author} {\bibfnamefont {Martin}\ \bibnamefont
  {Spinrath}},\ }\bibfield  {title} {\enquote {\bibinfo {title} {{Detection
  prospects for the Cosmic Neutrino Background using laser interferometers}},}\
  }\href {\doibase 10.1088/1475-7516/2017/06/055} {\bibfield  {journal}
  {\bibinfo  {journal} {JCAP}\ }\textbf {\bibinfo {volume} {06}},\ \bibinfo
  {pages} {055} (\bibinfo {year} {2017})},\ \Eprint
  {http://arxiv.org/abs/1703.08629} {arXiv:1703.08629 [astro-ph.CO]}
  \BibitemShut {NoStop}%
\bibitem [{\citenamefont {Alonso}\ \emph {et~al.}(2019)\citenamefont {Alonso},
  \citenamefont {Blas},\ and\ \citenamefont {Wolf}}]{Alonso:2018dxy}%
  \BibitemOpen
  \bibfield  {author} {\bibinfo {author} {\bibfnamefont {Rodrigo}\ \bibnamefont
  {Alonso}}, \bibinfo {author} {\bibfnamefont {Diego}\ \bibnamefont {Blas}}, \
  and\ \bibinfo {author} {\bibfnamefont {Peter}\ \bibnamefont {Wolf}},\
  }\bibfield  {title} {\enquote {\bibinfo {title} {{Exploring the ultra-light
  to sub-MeV dark matter window with atomic clocks and co-magnetometers}},}\
  }\href {\doibase 10.1007/JHEP07(2019)069} {\bibfield  {journal} {\bibinfo
  {journal} {JHEP}\ }\textbf {\bibinfo {volume} {07}},\ \bibinfo {pages} {069}
  (\bibinfo {year} {2019})},\ \Eprint {http://arxiv.org/abs/1810.00889}
  {arXiv:1810.00889 [hep-ph]} \BibitemShut {NoStop}%
\bibitem [{\citenamefont {Bauer}\ and\ \citenamefont
  {Shergold}(2023)}]{Bauer:2022lri}%
  \BibitemOpen
  \bibfield  {author} {\bibinfo {author} {\bibfnamefont {Martin}\ \bibnamefont
  {Bauer}}\ and\ \bibinfo {author} {\bibfnamefont {Jack~D.}\ \bibnamefont
  {Shergold}},\ }\bibfield  {title} {\enquote {\bibinfo {title} {{Limits on the
  cosmic neutrino background}},}\ }\href {\doibase
  10.1088/1475-7516/2023/01/003} {\bibfield  {journal} {\bibinfo  {journal}
  {JCAP}\ }\textbf {\bibinfo {volume} {01}},\ \bibinfo {pages} {003} (\bibinfo
  {year} {2023})},\ \Eprint {http://arxiv.org/abs/2207.12413} {arXiv:2207.12413
  [hep-ph]} \BibitemShut {NoStop}%
\bibitem [{\citenamefont {Arvanitaki}\ and\ \citenamefont
  {Dimopoulos}(2022)}]{Arvanitaki:2022oby}%
  \BibitemOpen
  \bibfield  {author} {\bibinfo {author} {\bibfnamefont {Asimina}\ \bibnamefont
  {Arvanitaki}}\ and\ \bibinfo {author} {\bibfnamefont {Savas}\ \bibnamefont
  {Dimopoulos}},\ }\bibfield  {title} {\enquote {\bibinfo {title} {{The Cosmic
  Neutrino Background on the Surface of the Earth}},}\ }\href@noop {} {\
  (\bibinfo {year} {2022})},\ \Eprint {http://arxiv.org/abs/2212.00036}
  {arXiv:2212.00036 [hep-ph]} \BibitemShut {NoStop}%
\bibitem [{\citenamefont {Aker}\ \emph {et~al.}(2022)\citenamefont {Aker} \emph
  {et~al.}}]{KATRIN:2021uub}%
  \BibitemOpen
  \bibfield  {author} {\bibinfo {author} {\bibfnamefont {M.}~\bibnamefont
  {Aker}} \emph {et~al.} (\bibinfo {collaboration} {KATRIN}),\ }\bibfield
  {title} {\enquote {\bibinfo {title} {{Direct neutrino-mass measurement with
  sub-electronvolt sensitivity}},}\ }\href {\doibase
  10.1038/s41567-021-01463-1} {\bibfield  {journal} {\bibinfo  {journal}
  {Nature Phys.}\ }\textbf {\bibinfo {volume} {18}},\ \bibinfo {pages}
  {160--166} (\bibinfo {year} {2022})},\ \Eprint
  {http://arxiv.org/abs/2105.08533} {arXiv:2105.08533 [hep-ex]} \BibitemShut
  {NoStop}%
\bibitem [{\citenamefont {Lesgourgues}\ and\ \citenamefont
  {Pastor}(2006)}]{Lesgourgues:2006nd}%
  \BibitemOpen
  \bibfield  {author} {\bibinfo {author} {\bibfnamefont {Julien}\ \bibnamefont
  {Lesgourgues}}\ and\ \bibinfo {author} {\bibfnamefont {Sergio}\ \bibnamefont
  {Pastor}},\ }\bibfield  {title} {\enquote {\bibinfo {title} {{Massive
  neutrinos and cosmology}},}\ }\href {\doibase 10.1016/j.physrep.2006.04.001}
  {\bibfield  {journal} {\bibinfo  {journal} {Phys. Rept.}\ }\textbf {\bibinfo
  {volume} {429}},\ \bibinfo {pages} {307--379} (\bibinfo {year} {2006})},\
  \Eprint {http://arxiv.org/abs/astro-ph/0603494} {arXiv:astro-ph/0603494}
  \BibitemShut {NoStop}%
\bibitem [{\citenamefont {Fields}\ \emph {et~al.}(2020)\citenamefont {Fields},
  \citenamefont {Olive}, \citenamefont {Yeh},\ and\ \citenamefont
  {Young}}]{Fields:2019pfx}%
  \BibitemOpen
  \bibfield  {author} {\bibinfo {author} {\bibfnamefont {Brian~D.}\
  \bibnamefont {Fields}}, \bibinfo {author} {\bibfnamefont {Keith~A.}\
  \bibnamefont {Olive}}, \bibinfo {author} {\bibfnamefont {Tsung-Han}\
  \bibnamefont {Yeh}}, \ and\ \bibinfo {author} {\bibfnamefont {Charles}\
  \bibnamefont {Young}},\ }\bibfield  {title} {\enquote {\bibinfo {title}
  {{Big-Bang Nucleosynthesis after Planck}},}\ }\href {\doibase
  10.1088/1475-7516/2020/03/010} {\bibfield  {journal} {\bibinfo  {journal}
  {JCAP}\ }\textbf {\bibinfo {volume} {03}},\ \bibinfo {pages} {010} (\bibinfo
  {year} {2020})},\ \bibinfo {note} {[Erratum: JCAP 11, E02 (2020)]},\ \Eprint
  {http://arxiv.org/abs/1912.01132} {arXiv:1912.01132 [astro-ph.CO]}
  \BibitemShut {NoStop}%
\bibitem [{\citenamefont {Aghanim}\ \emph {et~al.}(2020)\citenamefont {Aghanim}
  \emph {et~al.}}]{Planck:2018vyg}%
  \BibitemOpen
  \bibfield  {author} {\bibinfo {author} {\bibfnamefont {N.}~\bibnamefont
  {Aghanim}} \emph {et~al.} (\bibinfo {collaboration} {Planck}),\ }\bibfield
  {title} {\enquote {\bibinfo {title} {{Planck 2018 results. VI. Cosmological
  parameters}},}\ }\href {\doibase 10.1051/0004-6361/201833910} {\bibfield
  {journal} {\bibinfo  {journal} {Astron. Astrophys.}\ }\textbf {\bibinfo
  {volume} {641}},\ \bibinfo {pages} {A6} (\bibinfo {year} {2020})},\ \bibinfo
  {note} {[Erratum: Astron.Astrophys. 652, C4 (2021)]},\ \Eprint
  {http://arxiv.org/abs/1807.06209} {arXiv:1807.06209 [astro-ph.CO]}
  \BibitemShut {NoStop}%
\bibitem [{\citenamefont {Alvey}\ \emph
  {et~al.}(2022{\natexlab{b}})\citenamefont {Alvey}, \citenamefont {Escudero},\
  and\ \citenamefont {Sabti}}]{Alvey:2021sji}%
  \BibitemOpen
  \bibfield  {author} {\bibinfo {author} {\bibfnamefont {James}\ \bibnamefont
  {Alvey}}, \bibinfo {author} {\bibfnamefont {Miguel}\ \bibnamefont
  {Escudero}}, \ and\ \bibinfo {author} {\bibfnamefont {Nashwan}\ \bibnamefont
  {Sabti}},\ }\bibfield  {title} {\enquote {\bibinfo {title} {{What can CMB
  observations tell us about the neutrino distribution function?}}}\ }\href
  {\doibase 10.1088/1475-7516/2022/02/037} {\bibfield  {journal} {\bibinfo
  {journal} {JCAP}\ }\textbf {\bibinfo {volume} {02}},\ \bibinfo {pages} {037}
  (\bibinfo {year} {2022}{\natexlab{b}})},\ \Eprint
  {http://arxiv.org/abs/2111.12726} {arXiv:2111.12726 [astro-ph.CO]}
  \BibitemShut {NoStop}%
\bibitem [{\citenamefont {Esteban}\ \emph {et~al.}(2020)\citenamefont
  {Esteban}, \citenamefont {Gonzalez-Garcia}, \citenamefont {Maltoni},
  \citenamefont {Schwetz},\ and\ \citenamefont {Zhou}}]{Esteban:2020cvm}%
  \BibitemOpen
  \bibfield  {author} {\bibinfo {author} {\bibfnamefont {Ivan}\ \bibnamefont
  {Esteban}}, \bibinfo {author} {\bibfnamefont {M.~C.}\ \bibnamefont
  {Gonzalez-Garcia}}, \bibinfo {author} {\bibfnamefont {Michele}\ \bibnamefont
  {Maltoni}}, \bibinfo {author} {\bibfnamefont {Thomas}\ \bibnamefont
  {Schwetz}}, \ and\ \bibinfo {author} {\bibfnamefont {Albert}\ \bibnamefont
  {Zhou}},\ }\bibfield  {title} {\enquote {\bibinfo {title} {{The fate of
  hints: updated global analysis of three-flavor neutrino oscillations}},}\
  }\href {\doibase 10.1007/JHEP09(2020)178} {\bibfield  {journal} {\bibinfo
  {journal} {JHEP}\ }\textbf {\bibinfo {volume} {09}},\ \bibinfo {pages} {178}
  (\bibinfo {year} {2020})},\ \Eprint {http://arxiv.org/abs/2007.14792}
  {arXiv:2007.14792 [hep-ph]} \BibitemShut {NoStop}%
\bibitem [{\citenamefont {Ringwald}\ and\ \citenamefont
  {Wong}(2004)}]{Ringwald:2004np}%
  \BibitemOpen
  \bibfield  {author} {\bibinfo {author} {\bibfnamefont {Andreas}\ \bibnamefont
  {Ringwald}}\ and\ \bibinfo {author} {\bibfnamefont {Yvonne Y.~Y.}\
  \bibnamefont {Wong}},\ }\bibfield  {title} {\enquote {\bibinfo {title}
  {{Gravitational clustering of relic neutrinos and implications for their
  detection}},}\ }\href {\doibase 10.1088/1475-7516/2004/12/005} {\bibfield
  {journal} {\bibinfo  {journal} {JCAP}\ }\textbf {\bibinfo {volume} {12}},\
  \bibinfo {pages} {005} (\bibinfo {year} {2004})},\ \Eprint
  {http://arxiv.org/abs/hep-ph/0408241} {arXiv:hep-ph/0408241} \BibitemShut
  {NoStop}%
\bibitem [{\citenamefont {Villaescusa-Navarro}\ \emph
  {et~al.}(2011)\citenamefont {Villaescusa-Navarro}, \citenamefont
  {Miralda-Escud\'e}, \citenamefont {Pe\~na Garay},\ and\ \citenamefont
  {Quilis}}]{Villaescusa-Navarro:2011loy}%
  \BibitemOpen
  \bibfield  {author} {\bibinfo {author} {\bibfnamefont {Francisco}\
  \bibnamefont {Villaescusa-Navarro}}, \bibinfo {author} {\bibfnamefont
  {Jordi}\ \bibnamefont {Miralda-Escud\'e}}, \bibinfo {author} {\bibfnamefont
  {Carlos}\ \bibnamefont {Pe\~na Garay}}, \ and\ \bibinfo {author}
  {\bibfnamefont {Vicent}\ \bibnamefont {Quilis}},\ }\bibfield  {title}
  {\enquote {\bibinfo {title} {{Neutrino Halos in Clusters of Galaxies and
  their Weak Lensing Signature}},}\ }\href {\doibase
  10.1088/1475-7516/2011/06/027} {\bibfield  {journal} {\bibinfo  {journal}
  {JCAP}\ }\textbf {\bibinfo {volume} {06}},\ \bibinfo {pages} {027} (\bibinfo
  {year} {2011})},\ \Eprint {http://arxiv.org/abs/1104.4770} {arXiv:1104.4770
  [astro-ph.CO]} \BibitemShut {NoStop}%
\bibitem [{\citenamefont {Ichiki}\ and\ \citenamefont
  {Takada}(2012)}]{Ichiki:2011ue}%
  \BibitemOpen
  \bibfield  {author} {\bibinfo {author} {\bibfnamefont {Kiyotomo}\
  \bibnamefont {Ichiki}}\ and\ \bibinfo {author} {\bibfnamefont {Masahiro}\
  \bibnamefont {Takada}},\ }\bibfield  {title} {\enquote {\bibinfo {title}
  {{The impact of massive neutrinos on the abundance of massive clusters}},}\
  }\href {\doibase 10.1103/PhysRevD.85.063521} {\bibfield  {journal} {\bibinfo
  {journal} {Phys. Rev. D}\ }\textbf {\bibinfo {volume} {85}},\ \bibinfo
  {pages} {063521} (\bibinfo {year} {2012})},\ \Eprint
  {http://arxiv.org/abs/1108.4688} {arXiv:1108.4688 [astro-ph.CO]} \BibitemShut
  {NoStop}%
\bibitem [{\citenamefont {Zhang}\ and\ \citenamefont
  {Zhang}(2018)}]{Zhang:2017ljh}%
  \BibitemOpen
  \bibfield  {author} {\bibinfo {author} {\bibfnamefont {Jue}\ \bibnamefont
  {Zhang}}\ and\ \bibinfo {author} {\bibfnamefont {Xin}\ \bibnamefont
  {Zhang}},\ }\bibfield  {title} {\enquote {\bibinfo {title} {{Gravitational
  clustering of cosmic relic neutrinos in the Milky Way}},}\ }\href {\doibase
  10.1038/s41467-018-04264-y} {\bibfield  {journal} {\bibinfo  {journal}
  {Nature Commun.}\ }\textbf {\bibinfo {volume} {9}},\ \bibinfo {pages} {1833}
  (\bibinfo {year} {2018})},\ \Eprint {http://arxiv.org/abs/1712.01153}
  {arXiv:1712.01153 [astro-ph.CO]} \BibitemShut {NoStop}%
\bibitem [{\citenamefont {de~Salas}\ \emph {et~al.}(2017)\citenamefont
  {de~Salas}, \citenamefont {Gariazzo}, \citenamefont {Lesgourgues},\ and\
  \citenamefont {Pastor}}]{deSalas:2017wtt}%
  \BibitemOpen
  \bibfield  {author} {\bibinfo {author} {\bibfnamefont {P.~F.}\ \bibnamefont
  {de~Salas}}, \bibinfo {author} {\bibfnamefont {S.}~\bibnamefont {Gariazzo}},
  \bibinfo {author} {\bibfnamefont {J.}~\bibnamefont {Lesgourgues}}, \ and\
  \bibinfo {author} {\bibfnamefont {S.}~\bibnamefont {Pastor}},\ }\bibfield
  {title} {\enquote {\bibinfo {title} {{Calculation of the local density of
  relic neutrinos}},}\ }\href {\doibase 10.1088/1475-7516/2017/09/034}
  {\bibfield  {journal} {\bibinfo  {journal} {JCAP}\ }\textbf {\bibinfo
  {volume} {09}},\ \bibinfo {pages} {034} (\bibinfo {year} {2017})},\ \Eprint
  {http://arxiv.org/abs/1706.09850} {arXiv:1706.09850 [astro-ph.CO]}
  \BibitemShut {NoStop}%
\bibitem [{\citenamefont {Mertsch}\ \emph {et~al.}(2020)\citenamefont
  {Mertsch}, \citenamefont {Parimbelli}, \citenamefont {de~Salas},
  \citenamefont {Gariazzo}, \citenamefont {Lesgourgues},\ and\ \citenamefont
  {Pastor}}]{Mertsch:2019qjv}%
  \BibitemOpen
  \bibfield  {author} {\bibinfo {author} {\bibfnamefont {P.}~\bibnamefont
  {Mertsch}}, \bibinfo {author} {\bibfnamefont {G.}~\bibnamefont {Parimbelli}},
  \bibinfo {author} {\bibfnamefont {P.~F.}\ \bibnamefont {de~Salas}}, \bibinfo
  {author} {\bibfnamefont {S.}~\bibnamefont {Gariazzo}}, \bibinfo {author}
  {\bibfnamefont {J.}~\bibnamefont {Lesgourgues}}, \ and\ \bibinfo {author}
  {\bibfnamefont {S.}~\bibnamefont {Pastor}},\ }\bibfield  {title} {\enquote
  {\bibinfo {title} {{Neutrino clustering in the Milky Way and beyond}},}\
  }\href {\doibase 10.1088/1475-7516/2020/01/015} {\bibfield  {journal}
  {\bibinfo  {journal} {JCAP}\ }\textbf {\bibinfo {volume} {01}},\ \bibinfo
  {pages} {015} (\bibinfo {year} {2020})},\ \Eprint
  {http://arxiv.org/abs/1910.13388} {arXiv:1910.13388 [astro-ph.CO]}
  \BibitemShut {NoStop}%
\bibitem [{\citenamefont {Holm}\ \emph {et~al.}(2023)\citenamefont {Holm},
  \citenamefont {Oldengott},\ and\ \citenamefont {Zentarra}}]{Holm:2023rml}%
  \BibitemOpen
  \bibfield  {author} {\bibinfo {author} {\bibfnamefont {Emil~Brinch}\
  \bibnamefont {Holm}}, \bibinfo {author} {\bibfnamefont {Isabel~M.}\
  \bibnamefont {Oldengott}}, \ and\ \bibinfo {author} {\bibfnamefont {Stefan}\
  \bibnamefont {Zentarra}},\ }\bibfield  {title} {\enquote {\bibinfo {title}
  {{Local clustering of relic neutrinos with kinetic field theory}},}\
  }\href@noop {} {\  (\bibinfo {year} {2023})},\ \Eprint
  {http://arxiv.org/abs/2305.13379} {arXiv:2305.13379 [hep-ph]} \BibitemShut
  {NoStop}%
\bibitem [{\citenamefont {Seljak}\ and\ \citenamefont
  {Zaldarriaga}(2000)}]{Seljak:1999zn}%
  \BibitemOpen
  \bibfield  {author} {\bibinfo {author} {\bibfnamefont {Uros}\ \bibnamefont
  {Seljak}}\ and\ \bibinfo {author} {\bibfnamefont {Matias}\ \bibnamefont
  {Zaldarriaga}},\ }\bibfield  {title} {\enquote {\bibinfo {title} {{Lensing
  induced cluster signatures in cosmic microwave background}},}\ }\href
  {\doibase 10.1086/309098} {\bibfield  {journal} {\bibinfo  {journal}
  {Astrophys. J.}\ }\textbf {\bibinfo {volume} {538}},\ \bibinfo {pages}
  {57–64} (\bibinfo {year} {2000})},\ \Eprint
  {http://arxiv.org/abs/astro-ph/9907254} {arXiv:astro-ph/9907254} \BibitemShut
  {NoStop}%
\bibitem [{\citenamefont {Lewis}\ and\ \citenamefont
  {King}(2006)}]{Lewis:2005fq}%
  \BibitemOpen
  \bibfield  {author} {\bibinfo {author} {\bibfnamefont {Antony}\ \bibnamefont
  {Lewis}}\ and\ \bibinfo {author} {\bibfnamefont {Lindsay}\ \bibnamefont
  {King}},\ }\bibfield  {title} {\enquote {\bibinfo {title} {{Cluster masses
  from cmb and galaxy weak lensing}},}\ }\href {\doibase
  10.1103/PhysRevD.73.063006} {\bibfield  {journal} {\bibinfo  {journal} {Phys.
  Rev. D}\ }\textbf {\bibinfo {volume} {73}},\ \bibinfo {pages} {063006}
  (\bibinfo {year} {2006})},\ \Eprint {http://arxiv.org/abs/astro-ph/0512104}
  {arXiv:astro-ph/0512104} \BibitemShut {NoStop}%
\bibitem [{\citenamefont {Levy}\ \emph {et~al.}(2023)\citenamefont {Levy},
  \citenamefont {Raghunathan},\ and\ \citenamefont {Basu}}]{Levy:2023moy}%
  \BibitemOpen
  \bibfield  {author} {\bibinfo {author} {\bibfnamefont {Kevin}\ \bibnamefont
  {Levy}}, \bibinfo {author} {\bibfnamefont {Srinivasan}\ \bibnamefont
  {Raghunathan}}, \ and\ \bibinfo {author} {\bibfnamefont {Kaustuv}\
  \bibnamefont {Basu}},\ }\bibfield  {title} {\enquote {\bibinfo {title} {{A
  Foreground-Immune CMB-Cluster Lensing Estimator}},}\ }\href@noop {} {\
  (\bibinfo {year} {2023})},\ \Eprint {http://arxiv.org/abs/2305.06326}
  {arXiv:2305.06326 [astro-ph.CO]} \BibitemShut {NoStop}%
\bibitem [{\citenamefont {Horowitz}\ \emph {et~al.}(2019)\citenamefont
  {Horowitz}, \citenamefont {Ferraro},\ and\ \citenamefont
  {Sherwin}}]{Horowitz:2017iql}%
  \BibitemOpen
  \bibfield  {author} {\bibinfo {author} {\bibfnamefont {Benjamin}\
  \bibnamefont {Horowitz}}, \bibinfo {author} {\bibfnamefont {Simone}\
  \bibnamefont {Ferraro}}, \ and\ \bibinfo {author} {\bibfnamefont {Blake~D.}\
  \bibnamefont {Sherwin}},\ }\bibfield  {title} {\enquote {\bibinfo {title}
  {{Reconstructing Small Scale Lenses from the Cosmic Microwave Background
  Temperature Fluctuations}},}\ }\href {\doibase 10.1093/mnras/stz566}
  {\bibfield  {journal} {\bibinfo  {journal} {Mon. Not. Roy. Astron. Soc.}\
  }\textbf {\bibinfo {volume} {485}},\ \bibinfo {pages} {3919--3929} (\bibinfo
  {year} {2019})},\ \Eprint {http://arxiv.org/abs/1710.10236} {arXiv:1710.10236
  [astro-ph.CO]} \BibitemShut {NoStop}%
\bibitem [{\citenamefont {Birkinshaw}\ and\ \citenamefont
  {Gull}(1983)}]{Birkinshaw:1983}%
  \BibitemOpen
  \bibfield  {author} {\bibinfo {author} {\bibfnamefont {M.}~\bibnamefont
  {Birkinshaw}}\ and\ \bibinfo {author} {\bibfnamefont {S.~F.}\ \bibnamefont
  {Gull}},\ }\bibfield  {title} {\enquote {\bibinfo {title} {{A test for
  transverse motions of galaxies and clusters}},}\ }\href@noop {} {\bibfield
  {journal} {\bibinfo  {journal} {Nature}\ }\textbf {\bibinfo {volume} {302}},\
  \bibinfo {pages} {315--317} (\bibinfo {year} {1983})}\BibitemShut {NoStop}%
\bibitem [{\citenamefont {{Hotinli}}\ \emph {et~al.}(2019)\citenamefont
  {{Hotinli}}, \citenamefont {{Meyers}}, \citenamefont {{Dalal}}, \citenamefont
  {{Jaffe}}, \citenamefont {{Johnson}}, \citenamefont {{Mertens}},
  \citenamefont {{M{\"u}nchmeyer}}, \citenamefont {{Smith}},\ and\
  \citenamefont {{van Engelen}}}]{Hotinli:2018yyc}%
  \BibitemOpen
  \bibfield  {author} {\bibinfo {author} {\bibfnamefont {Selim~C.}\
  \bibnamefont {{Hotinli}}}, \bibinfo {author} {\bibfnamefont {Joel}\
  \bibnamefont {{Meyers}}}, \bibinfo {author} {\bibfnamefont {Neal}\
  \bibnamefont {{Dalal}}}, \bibinfo {author} {\bibfnamefont {Andrew~H.}\
  \bibnamefont {{Jaffe}}}, \bibinfo {author} {\bibfnamefont {Matthew~C.}\
  \bibnamefont {{Johnson}}}, \bibinfo {author} {\bibfnamefont {James~B.}\
  \bibnamefont {{Mertens}}}, \bibinfo {author} {\bibfnamefont {Moritz}\
  \bibnamefont {{M{\"u}nchmeyer}}}, \bibinfo {author} {\bibfnamefont
  {Kendrick~M.}\ \bibnamefont {{Smith}}}, \ and\ \bibinfo {author}
  {\bibfnamefont {Alexander}\ \bibnamefont {{van Engelen}}},\ }\bibfield
  {title} {\enquote {\bibinfo {title} {{Transverse Velocities with the Moving
  Lens Effect}},}\ }\href {\doibase 10.1103/PhysRevLett.123.061301} {\bibfield
  {journal} {\bibinfo  {journal} {\prl}\ }\textbf {\bibinfo {volume} {123}},\
  \bibinfo {eid} {061301} (\bibinfo {year} {2019})},\ \Eprint
  {http://arxiv.org/abs/1812.03167} {arXiv:1812.03167 [astro-ph.CO]}
  \BibitemShut {NoStop}%
\bibitem [{\citenamefont {{Hotinli}}\ \emph {et~al.}(2021)\citenamefont
  {{Hotinli}}, \citenamefont {{Johnson}},\ and\ \citenamefont
  {{Meyers}}}]{Hotinli:2020ntd}%
  \BibitemOpen
  \bibfield  {author} {\bibinfo {author} {\bibfnamefont {Selim~C.}\
  \bibnamefont {{Hotinli}}}, \bibinfo {author} {\bibfnamefont {Matthew~C.}\
  \bibnamefont {{Johnson}}}, \ and\ \bibinfo {author} {\bibfnamefont {Joel}\
  \bibnamefont {{Meyers}}},\ }\bibfield  {title} {\enquote {\bibinfo {title}
  {{Optimal filters for the moving lens effect}},}\ }\href {\doibase
  10.1103/PhysRevD.103.043536} {\bibfield  {journal} {\bibinfo  {journal}
  {\prd}\ }\textbf {\bibinfo {volume} {103}},\ \bibinfo {eid} {043536}
  (\bibinfo {year} {2021})},\ \Eprint {http://arxiv.org/abs/2006.03060}
  {arXiv:2006.03060 [astro-ph.CO]} \BibitemShut {NoStop}%
\bibitem [{\citenamefont {Correa}\ \emph {et~al.}(2015)\citenamefont {Correa},
  \citenamefont {Wyithe}, \citenamefont {Schaye},\ and\ \citenamefont
  {Duffy}}]{Correa:2015dva}%
  \BibitemOpen
  \bibfield  {author} {\bibinfo {author} {\bibfnamefont {Camila~A.}\
  \bibnamefont {Correa}}, \bibinfo {author} {\bibfnamefont {J.~Stuart~B.}\
  \bibnamefont {Wyithe}}, \bibinfo {author} {\bibfnamefont {Joop}\ \bibnamefont
  {Schaye}}, \ and\ \bibinfo {author} {\bibfnamefont {Alan~R.}\ \bibnamefont
  {Duffy}},\ }\bibfield  {title} {\enquote {\bibinfo {title} {{The accretion
  history of dark matter haloes \textendash{} III. A physical model for the
  concentration\textendash{}mass relation}},}\ }\href {\doibase
  10.1093/mnras/stv1363} {\bibfield  {journal} {\bibinfo  {journal} {Mon. Not.
  Roy. Astron. Soc.}\ }\textbf {\bibinfo {volume} {452}},\ \bibinfo {pages}
  {1217--1232} (\bibinfo {year} {2015})},\ \Eprint
  {http://arxiv.org/abs/1502.00391} {arXiv:1502.00391 [astro-ph.CO]}
  \BibitemShut {NoStop}%
\bibitem [{\citenamefont {Singh}\ and\ \citenamefont
  {Ma}(2003)}]{Singh:2002de}%
  \BibitemOpen
  \bibfield  {author} {\bibinfo {author} {\bibfnamefont {Shwetabh}\
  \bibnamefont {Singh}}\ and\ \bibinfo {author} {\bibfnamefont {Chung-Pei}\
  \bibnamefont {Ma}},\ }\bibfield  {title} {\enquote {\bibinfo {title}
  {{Neutrino clustering in cold dark matter halos : Implications for
  ultrahigh-energy cosmic rays}},}\ }\href {\doibase
  10.1103/PhysRevD.67.023506} {\bibfield  {journal} {\bibinfo  {journal} {Phys.
  Rev. D}\ }\textbf {\bibinfo {volume} {67}},\ \bibinfo {pages} {023506}
  (\bibinfo {year} {2003})},\ \Eprint {http://arxiv.org/abs/astro-ph/0208419}
  {arXiv:astro-ph/0208419} \BibitemShut {NoStop}%
\bibitem [{\citenamefont {Blas}\ \emph {et~al.}(2011)\citenamefont {Blas},
  \citenamefont {Lesgourgues},\ and\ \citenamefont {Tram}}]{Blas:2011rf}%
  \BibitemOpen
  \bibfield  {author} {\bibinfo {author} {\bibfnamefont {Diego}\ \bibnamefont
  {Blas}}, \bibinfo {author} {\bibfnamefont {Julien}\ \bibnamefont
  {Lesgourgues}}, \ and\ \bibinfo {author} {\bibfnamefont {Thomas}\
  \bibnamefont {Tram}},\ }\bibfield  {title} {\enquote {\bibinfo {title} {{The
  Cosmic Linear Anisotropy Solving System (CLASS) II: Approximation
  schemes}},}\ }\href {\doibase 10.1088/1475-7516/2011/07/034} {\bibfield
  {journal} {\bibinfo  {journal} {JCAP}\ }\textbf {\bibinfo {volume} {07}},\
  \bibinfo {pages} {034} (\bibinfo {year} {2011})},\ \Eprint
  {http://arxiv.org/abs/1104.2933} {arXiv:1104.2933 [astro-ph.CO]} \BibitemShut
  {NoStop}%
\bibitem [{\citenamefont {Lesgourgues}\ and\ \citenamefont
  {Pastor}(2012)}]{Lesgourgues:2012uu}%
  \BibitemOpen
  \bibfield  {author} {\bibinfo {author} {\bibfnamefont {Julien}\ \bibnamefont
  {Lesgourgues}}\ and\ \bibinfo {author} {\bibfnamefont {Sergio}\ \bibnamefont
  {Pastor}},\ }\bibfield  {title} {\enquote {\bibinfo {title} {{Neutrino mass
  from Cosmology}},}\ }\href {\doibase 10.1155/2012/608515} {\bibfield
  {journal} {\bibinfo  {journal} {Adv. High Energy Phys.}\ }\textbf {\bibinfo
  {volume} {2012}},\ \bibinfo {pages} {608515} (\bibinfo {year} {2012})},\
  \Eprint {http://arxiv.org/abs/1212.6154} {arXiv:1212.6154 [hep-ph]}
  \BibitemShut {NoStop}%
\bibitem [{\citenamefont {{Lewis}}\ and\ \citenamefont
  {{Challinor}}(2006)}]{Lewis:2006fu}%
  \BibitemOpen
  \bibfield  {author} {\bibinfo {author} {\bibfnamefont {Antony}\ \bibnamefont
  {{Lewis}}}\ and\ \bibinfo {author} {\bibfnamefont {Anthony}\ \bibnamefont
  {{Challinor}}},\ }\bibfield  {title} {\enquote {\bibinfo {title} {{Weak
  gravitational lensing of the CMB}},}\ }\href {\doibase
  10.1016/j.physrep.2006.03.002} {\bibfield  {journal} {\bibinfo  {journal}
  {Physics Reports}\ }\textbf {\bibinfo {volume} {429}},\ \bibinfo {pages}
  {1--65} (\bibinfo {year} {2006})},\ \Eprint
  {http://arxiv.org/abs/astro-ph/0601594} {arXiv:astro-ph/0601594 [astro-ph]}
  \BibitemShut {NoStop}%
\bibitem [{\citenamefont {Ferraro}\ \emph {et~al.}(2022)\citenamefont
  {Ferraro}, \citenamefont {Schaan},\ and\ \citenamefont
  {Pierpaoli}}]{Ferraro:2022twg}%
  \BibitemOpen
  \bibfield  {author} {\bibinfo {author} {\bibfnamefont {Simone}\ \bibnamefont
  {Ferraro}}, \bibinfo {author} {\bibfnamefont {Emmanuel}\ \bibnamefont
  {Schaan}}, \ and\ \bibinfo {author} {\bibfnamefont {Elena}\ \bibnamefont
  {Pierpaoli}},\ }\bibfield  {title} {\enquote {\bibinfo {title} {{Is the
  Rees-Sciama effect detectable by the next generation of cosmological
  experiments?}}}\ }\href@noop {} {\  (\bibinfo {year} {2022})},\ \Eprint
  {http://arxiv.org/abs/2205.10332} {arXiv:2205.10332 [astro-ph.CO]}
  \BibitemShut {NoStop}%
\bibitem [{\citenamefont {Abell}\ \emph {et~al.}(2009)\citenamefont {Abell}
  \emph {et~al.}}]{LSSTScience:2009jmu}%
  \BibitemOpen
  \bibfield  {author} {\bibinfo {author} {\bibfnamefont {Paul~A.}\ \bibnamefont
  {Abell}} \emph {et~al.} (\bibinfo {collaboration} {LSST Science, LSST
  Project}),\ }\bibfield  {title} {\enquote {\bibinfo {title} {{LSST Science
  Book, Version 2.0}},}\ }\href@noop {} {\  (\bibinfo {year} {2009})},\ \Eprint
  {http://arxiv.org/abs/0912.0201} {arXiv:0912.0201 [astro-ph.IM]} \BibitemShut
  {NoStop}%
\bibitem [{\citenamefont {Sheth}\ and\ \citenamefont
  {Tormen}(1999)}]{Sheth:1999mn}%
  \BibitemOpen
  \bibfield  {author} {\bibinfo {author} {\bibfnamefont {Ravi~K.}\ \bibnamefont
  {Sheth}}\ and\ \bibinfo {author} {\bibfnamefont {Giuseppe}\ \bibnamefont
  {Tormen}},\ }\bibfield  {title} {\enquote {\bibinfo {title} {{Large scale
  bias and the peak background split}},}\ }\href {\doibase
  10.1046/j.1365-8711.1999.02692.x} {\bibfield  {journal} {\bibinfo  {journal}
  {Mon. Not. Roy. Astron. Soc.}\ }\textbf {\bibinfo {volume} {308}},\ \bibinfo
  {pages} {119} (\bibinfo {year} {1999})},\ \Eprint
  {http://arxiv.org/abs/astro-ph/9901122} {arXiv:astro-ph/9901122} \BibitemShut
  {NoStop}%
\bibitem [{\citenamefont {Palmese}\ \emph {et~al.}(2020)\citenamefont {Palmese}
  \emph {et~al.}}]{Palmese:2019lkh}%
  \BibitemOpen
  \bibfield  {author} {\bibinfo {author} {\bibfnamefont {A.}~\bibnamefont
  {Palmese}} \emph {et~al.} (\bibinfo {collaboration} {DES}),\ }\bibfield
  {title} {\enquote {\bibinfo {title} {{Stellar Mass as a Galaxy Cluster Mass
  Proxy: Application to the Dark Energy Survey redMaPPer Clusters}},}\ }\href
  {\doibase 10.1093/mnras/staa526} {\bibfield  {journal} {\bibinfo  {journal}
  {Mon. Not. Roy. Astron. Soc.}\ }\textbf {\bibinfo {volume} {493}},\ \bibinfo
  {pages} {4591--4606} (\bibinfo {year} {2020})},\ \Eprint
  {http://arxiv.org/abs/1903.08813} {arXiv:1903.08813 [astro-ph.CO]}
  \BibitemShut {NoStop}%
\bibitem [{\citenamefont {Murata}\ \emph {et~al.}(2018)\citenamefont {Murata},
  \citenamefont {Nishimichi}, \citenamefont {Takada}, \citenamefont {Miyatake},
  \citenamefont {Shirasaki}, \citenamefont {More}, \citenamefont {Takahashi},\
  and\ \citenamefont {Osato}}]{Murata:2017zdo}%
  \BibitemOpen
  \bibfield  {author} {\bibinfo {author} {\bibfnamefont {Ryoma}\ \bibnamefont
  {Murata}}, \bibinfo {author} {\bibfnamefont {Takahiro}\ \bibnamefont
  {Nishimichi}}, \bibinfo {author} {\bibfnamefont {Masahiro}\ \bibnamefont
  {Takada}}, \bibinfo {author} {\bibfnamefont {Hironao}\ \bibnamefont
  {Miyatake}}, \bibinfo {author} {\bibfnamefont {Masato}\ \bibnamefont
  {Shirasaki}}, \bibinfo {author} {\bibfnamefont {Surhud}\ \bibnamefont
  {More}}, \bibinfo {author} {\bibfnamefont {Ryuichi}\ \bibnamefont
  {Takahashi}}, \ and\ \bibinfo {author} {\bibfnamefont {Ken}\ \bibnamefont
  {Osato}},\ }\bibfield  {title} {\enquote {\bibinfo {title} {{Constraints on
  the mass-richness relation from the abundance and weak lensing of SDSS
  clusters}},}\ }\href {\doibase 10.3847/1538-4357/aaaab8} {\bibfield
  {journal} {\bibinfo  {journal} {Astrophys. J.}\ }\textbf {\bibinfo {volume}
  {854}},\ \bibinfo {pages} {120} (\bibinfo {year} {2018})},\ \Eprint
  {http://arxiv.org/abs/1707.01907} {arXiv:1707.01907 [astro-ph.CO]}
  \BibitemShut {NoStop}%
\bibitem [{\citenamefont {Ballardini}\ \emph {et~al.}(2019)\citenamefont
  {Ballardini}, \citenamefont {Matthewson},\ and\ \citenamefont
  {Maartens}}]{Ballardini:2019wxj}%
  \BibitemOpen
  \bibfield  {author} {\bibinfo {author} {\bibfnamefont {Mario}\ \bibnamefont
  {Ballardini}}, \bibinfo {author} {\bibfnamefont {William~Luke}\ \bibnamefont
  {Matthewson}}, \ and\ \bibinfo {author} {\bibfnamefont {Roy}\ \bibnamefont
  {Maartens}},\ }\bibfield  {title} {\enquote {\bibinfo {title} {{Constraining
  primordial non-Gaussianity using two galaxy surveys and CMB lensing}},}\
  }\href {\doibase 10.1093/mnras/stz2258} {\bibfield  {journal} {\bibinfo
  {journal} {Mon. Not. Roy. Astron. Soc.}\ }\textbf {\bibinfo {volume} {489}},\
  \bibinfo {pages} {1950--1956} (\bibinfo {year} {2019})},\ \Eprint
  {http://arxiv.org/abs/1906.04730} {arXiv:1906.04730 [astro-ph.CO]}
  \BibitemShut {NoStop}%
\bibitem [{\citenamefont {Schlegel}\ \emph {et~al.}(2019)\citenamefont
  {Schlegel} \emph {et~al.}}]{Schlegel:2019eqc}%
  \BibitemOpen
  \bibfield  {author} {\bibinfo {author} {\bibfnamefont {David~J.}\
  \bibnamefont {Schlegel}} \emph {et~al.},\ }\bibfield  {title} {\enquote
  {\bibinfo {title} {{Astro2020 APC White Paper: The MegaMapper: a z
  \ensuremath{>} 2 Spectroscopic Instrument for the Study of Inflation and Dark
  Energy}},}\ }\href@noop {} {\  (\bibinfo {year} {2019})},\ \Eprint
  {http://arxiv.org/abs/1907.11171} {arXiv:1907.11171 [astro-ph.IM]}
  \BibitemShut {NoStop}%
\bibitem [{\citenamefont {{Lewis}}\ \emph {et~al.}(2000)\citenamefont
  {{Lewis}}, \citenamefont {{Challinor}},\ and\ \citenamefont
  {{Lasenby}}}]{CAMB}%
  \BibitemOpen
  \bibfield  {author} {\bibinfo {author} {\bibfnamefont {Antony}\ \bibnamefont
  {{Lewis}}}, \bibinfo {author} {\bibfnamefont {Anthony}\ \bibnamefont
  {{Challinor}}}, \ and\ \bibinfo {author} {\bibfnamefont {Anthony}\
  \bibnamefont {{Lasenby}}},\ }\bibfield  {title} {\enquote {\bibinfo {title}
  {{Efficient Computation of Cosmic Microwave Background Anisotropies in Closed
  Friedmann-Robertson-Walker Models}},}\ }\href {\doibase 10.1086/309179}
  {\bibfield  {journal} {\bibinfo  {journal} {Astrophys. J.}\ }\textbf
  {\bibinfo {volume} {538}},\ \bibinfo {pages} {473--476} (\bibinfo {year}
  {2000})},\ \Eprint {http://arxiv.org/abs/astro-ph/9911177}
  {arXiv:astro-ph/9911177 [astro-ph]} \BibitemShut {NoStop}%
\bibitem [{\citenamefont {Hotinli}\ \emph {et~al.}(2021)\citenamefont
  {Hotinli}, \citenamefont {Smith}, \citenamefont {Madhavacheril},\ and\
  \citenamefont {Kamionkowski}}]{Hotinli:2021hih}%
  \BibitemOpen
  \bibfield  {author} {\bibinfo {author} {\bibfnamefont {Selim~C.}\
  \bibnamefont {Hotinli}}, \bibinfo {author} {\bibfnamefont {Kendrick~M.}\
  \bibnamefont {Smith}}, \bibinfo {author} {\bibfnamefont {Mathew~S.}\
  \bibnamefont {Madhavacheril}}, \ and\ \bibinfo {author} {\bibfnamefont
  {Marc}\ \bibnamefont {Kamionkowski}},\ }\bibfield  {title} {\enquote
  {\bibinfo {title} {{Cosmology with the moving lens effect}},}\ }\href
  {\doibase 10.1103/PhysRevD.104.083529} {\bibfield  {journal} {\bibinfo
  {journal} {Phys. Rev. D}\ }\textbf {\bibinfo {volume} {104}},\ \bibinfo
  {pages} {083529} (\bibinfo {year} {2021})},\ \Eprint
  {http://arxiv.org/abs/2108.02207} {arXiv:2108.02207 [astro-ph.CO]}
  \BibitemShut {NoStop}%
\bibitem [{\citenamefont {Abitbol}\ \emph {et~al.}(2019)\citenamefont {Abitbol}
  \emph {et~al.}}]{SimonsObservatory:2019qwx}%
  \BibitemOpen
  \bibfield  {author} {\bibinfo {author} {\bibfnamefont {Maximilian~H.}\
  \bibnamefont {Abitbol}} \emph {et~al.} (\bibinfo {collaboration} {Simons
  Observatory}),\ }\bibfield  {title} {\enquote {\bibinfo {title} {{The Simons
  Observatory: Astro2020 Decadal Project Whitepaper}},}\ }\href@noop {}
  {\bibfield  {journal} {\bibinfo  {journal} {Bull. Am. Astron. Soc.}\ }\textbf
  {\bibinfo {volume} {51}},\ \bibinfo {pages} {147} (\bibinfo {year} {2019})},\
  \Eprint {http://arxiv.org/abs/1907.08284} {arXiv:1907.08284 [astro-ph.IM]}
  \BibitemShut {NoStop}%
\bibitem [{\citenamefont {Abazajian}\ \emph {et~al.}(2016)\citenamefont
  {Abazajian} \emph {et~al.}}]{CMB-S4:2016ple}%
  \BibitemOpen
  \bibfield  {author} {\bibinfo {author} {\bibfnamefont {Kevork~N.}\
  \bibnamefont {Abazajian}} \emph {et~al.} (\bibinfo {collaboration}
  {CMB-S4}),\ }\bibfield  {title} {\enquote {\bibinfo {title} {{CMB-S4 Science
  Book, First Edition}},}\ }\href@noop {} {\  (\bibinfo {year} {2016})},\
  \Eprint {http://arxiv.org/abs/1610.02743} {arXiv:1610.02743 [astro-ph.CO]}
  \BibitemShut {NoStop}%
\bibitem [{\citenamefont {Abazajian}\ \emph {et~al.}(2019)\citenamefont
  {Abazajian} \emph {et~al.}}]{Abazajian:2019eic}%
  \BibitemOpen
  \bibfield  {author} {\bibinfo {author} {\bibfnamefont {Kevork}\ \bibnamefont
  {Abazajian}} \emph {et~al.},\ }\bibfield  {title} {\enquote {\bibinfo {title}
  {{CMB-S4 Science Case, Reference Design, and Project Plan}},}\ }\href@noop {}
  {\  (\bibinfo {year} {2019})},\ \Eprint {http://arxiv.org/abs/1907.04473}
  {arXiv:1907.04473 [astro-ph.IM]} \BibitemShut {NoStop}%
\bibitem [{\citenamefont {Sehgal}\ \emph {et~al.}(2019)\citenamefont {Sehgal}
  \emph {et~al.}}]{Sehgal:2019ewc}%
  \BibitemOpen
  \bibfield  {author} {\bibinfo {author} {\bibfnamefont {Neelima}\ \bibnamefont
  {Sehgal}} \emph {et~al.},\ }\bibfield  {title} {\enquote {\bibinfo {title}
  {{CMB-HD: An Ultra-Deep, High-Resolution Millimeter-Wave Survey Over Half the
  Sky}},}\ }\href@noop {} {\  (\bibinfo {year} {2019})},\ \Eprint
  {http://arxiv.org/abs/1906.10134} {arXiv:1906.10134 [astro-ph.CO]}
  \BibitemShut {NoStop}%
\bibitem [{\citenamefont {Aiola}\ \emph {et~al.}(2022)\citenamefont {Aiola}
  \emph {et~al.}}]{CMB-HD:2022bsz}%
  \BibitemOpen
  \bibfield  {author} {\bibinfo {author} {\bibfnamefont {Simone}\ \bibnamefont
  {Aiola}} \emph {et~al.} (\bibinfo {collaboration} {CMB-HD}),\ }\bibfield
  {title} {\enquote {\bibinfo {title} {{Snowmass2021 CMB-HD White Paper}},}\
  }\href@noop {} {\  (\bibinfo {year} {2022})},\ \Eprint
  {http://arxiv.org/abs/2203.05728} {arXiv:2203.05728 [astro-ph.CO]}
  \BibitemShut {NoStop}%
\bibitem [{\citenamefont {Namikawa}\ \emph {et~al.}(2013)\citenamefont
  {Namikawa}, \citenamefont {Hanson},\ and\ \citenamefont
  {Takahashi}}]{Namikawa:2012pe}%
  \BibitemOpen
  \bibfield  {author} {\bibinfo {author} {\bibfnamefont {Toshiya}\ \bibnamefont
  {Namikawa}}, \bibinfo {author} {\bibfnamefont {Duncan}\ \bibnamefont
  {Hanson}}, \ and\ \bibinfo {author} {\bibfnamefont {Ryuichi}\ \bibnamefont
  {Takahashi}},\ }\bibfield  {title} {\enquote {\bibinfo {title}
  {{Bias-Hardened CMB Lensing}},}\ }\href {\doibase 10.1093/mnras/stt195}
  {\bibfield  {journal} {\bibinfo  {journal} {Mon. Not. Roy. Astron. Soc.}\
  }\textbf {\bibinfo {volume} {431}},\ \bibinfo {pages} {609--620} (\bibinfo
  {year} {2013})},\ \Eprint {http://arxiv.org/abs/1209.0091} {arXiv:1209.0091
  [astro-ph.CO]} \BibitemShut {NoStop}%
\bibitem [{\citenamefont {Millea}\ \emph {et~al.}(2019)\citenamefont {Millea},
  \citenamefont {Anderes},\ and\ \citenamefont {Wandelt}}]{Millea:2017fyd}%
  \BibitemOpen
  \bibfield  {author} {\bibinfo {author} {\bibfnamefont {Marius}\ \bibnamefont
  {Millea}}, \bibinfo {author} {\bibfnamefont {Ethan}\ \bibnamefont {Anderes}},
  \ and\ \bibinfo {author} {\bibfnamefont {Benjamin~D.}\ \bibnamefont
  {Wandelt}},\ }\bibfield  {title} {\enquote {\bibinfo {title} {{Bayesian
  delensing of CMB temperature and polarization}},}\ }\href {\doibase
  10.1103/PhysRevD.100.023509} {\bibfield  {journal} {\bibinfo  {journal}
  {Phys. Rev. D}\ }\textbf {\bibinfo {volume} {100}},\ \bibinfo {pages}
  {023509} (\bibinfo {year} {2019})},\ \Eprint
  {http://arxiv.org/abs/1708.06753} {arXiv:1708.06753 [astro-ph.CO]}
  \BibitemShut {NoStop}%
\bibitem [{\citenamefont {Hadzhiyska}\ \emph {et~al.}(2019)\citenamefont
  {Hadzhiyska}, \citenamefont {Sherwin}, \citenamefont {Madhavacheril},\ and\
  \citenamefont {Ferraro}}]{Hadzhiyska:2019cle}%
  \BibitemOpen
  \bibfield  {author} {\bibinfo {author} {\bibfnamefont {Boryana}\ \bibnamefont
  {Hadzhiyska}}, \bibinfo {author} {\bibfnamefont {Blake~D.}\ \bibnamefont
  {Sherwin}}, \bibinfo {author} {\bibfnamefont {Mathew}\ \bibnamefont
  {Madhavacheril}}, \ and\ \bibinfo {author} {\bibfnamefont {Simone}\
  \bibnamefont {Ferraro}},\ }\bibfield  {title} {\enquote {\bibinfo {title}
  {{Improving Small-Scale CMB Lensing Reconstruction}},}\ }\href {\doibase
  10.1103/PhysRevD.100.023547} {\bibfield  {journal} {\bibinfo  {journal}
  {Phys. Rev. D}\ }\textbf {\bibinfo {volume} {100}},\ \bibinfo {pages}
  {023547} (\bibinfo {year} {2019})},\ \Eprint
  {http://arxiv.org/abs/1905.04217} {arXiv:1905.04217 [astro-ph.CO]}
  \BibitemShut {NoStop}%
\bibitem [{\citenamefont {Darwish}\ \emph {et~al.}(2023)\citenamefont
  {Darwish}, \citenamefont {Sherwin}, \citenamefont {Sailer}, \citenamefont
  {Schaan},\ and\ \citenamefont {Ferraro}}]{Darwish:2021ycf}%
  \BibitemOpen
  \bibfield  {author} {\bibinfo {author} {\bibfnamefont {Omar}\ \bibnamefont
  {Darwish}}, \bibinfo {author} {\bibfnamefont {Blake~D.}\ \bibnamefont
  {Sherwin}}, \bibinfo {author} {\bibfnamefont {Noah}\ \bibnamefont {Sailer}},
  \bibinfo {author} {\bibfnamefont {Emmanuel}\ \bibnamefont {Schaan}}, \ and\
  \bibinfo {author} {\bibfnamefont {Simone}\ \bibnamefont {Ferraro}},\
  }\bibfield  {title} {\enquote {\bibinfo {title} {{Optimizing foreground
  mitigation for CMB lensing with combined multifrequency and geometric
  methods}},}\ }\href {\doibase 10.1103/PhysRevD.107.043519} {\bibfield
  {journal} {\bibinfo  {journal} {Phys. Rev. D}\ }\textbf {\bibinfo {volume}
  {107}},\ \bibinfo {pages} {043519} (\bibinfo {year} {2023})},\ \Eprint
  {http://arxiv.org/abs/2111.00462} {arXiv:2111.00462 [astro-ph.CO]}
  \BibitemShut {NoStop}%
\bibitem [{\citenamefont {Sailer}\ \emph {et~al.}(2020)\citenamefont {Sailer},
  \citenamefont {Schaan},\ and\ \citenamefont {Ferraro}}]{Sailer:2020lal}%
  \BibitemOpen
  \bibfield  {author} {\bibinfo {author} {\bibfnamefont {Noah}\ \bibnamefont
  {Sailer}}, \bibinfo {author} {\bibfnamefont {Emmanuel}\ \bibnamefont
  {Schaan}}, \ and\ \bibinfo {author} {\bibfnamefont {Simone}\ \bibnamefont
  {Ferraro}},\ }\bibfield  {title} {\enquote {\bibinfo {title} {{Lower bias,
  lower noise CMB lensing with foreground-hardened estimators}},}\ }\href
  {\doibase 10.1103/PhysRevD.102.063517} {\bibfield  {journal} {\bibinfo
  {journal} {Phys. Rev. D}\ }\textbf {\bibinfo {volume} {102}},\ \bibinfo
  {pages} {063517} (\bibinfo {year} {2020})},\ \Eprint
  {http://arxiv.org/abs/2007.04325} {arXiv:2007.04325 [astro-ph.CO]}
  \BibitemShut {NoStop}%
\bibitem [{\citenamefont {Sailer}\ \emph {et~al.}(2023)\citenamefont {Sailer},
  \citenamefont {Ferraro},\ and\ \citenamefont {Schaan}}]{Sailer:2022jwt}%
  \BibitemOpen
  \bibfield  {author} {\bibinfo {author} {\bibfnamefont {Noah}\ \bibnamefont
  {Sailer}}, \bibinfo {author} {\bibfnamefont {Simone}\ \bibnamefont
  {Ferraro}}, \ and\ \bibinfo {author} {\bibfnamefont {Emmanuel}\ \bibnamefont
  {Schaan}},\ }\bibfield  {title} {\enquote {\bibinfo {title}
  {{Foreground-immune CMB lensing reconstruction with polarization}},}\ }\href
  {\doibase 10.1103/PhysRevD.107.023504} {\bibfield  {journal} {\bibinfo
  {journal} {Phys. Rev. D}\ }\textbf {\bibinfo {volume} {107}},\ \bibinfo
  {pages} {023504} (\bibinfo {year} {2023})},\ \Eprint
  {http://arxiv.org/abs/2211.03786} {arXiv:2211.03786 [astro-ph.CO]}
  \BibitemShut {NoStop}%
\bibitem [{\citenamefont {{Parker}}\ \emph {et~al.}(2022)\citenamefont
  {{Parker}}, \citenamefont {{Han}}, \citenamefont {{Lemos Portela}},\ and\
  \citenamefont {{Ho}}}]{2022arXiv221108990P}%
  \BibitemOpen
  \bibfield  {author} {\bibinfo {author} {\bibfnamefont {Liam}\ \bibnamefont
  {{Parker}}}, \bibinfo {author} {\bibfnamefont {Dongwon}\ \bibnamefont
  {{Han}}}, \bibinfo {author} {\bibfnamefont {Pablo}\ \bibnamefont {{Lemos
  Portela}}}, \ and\ \bibinfo {author} {\bibfnamefont {Shirley}\ \bibnamefont
  {{Ho}}},\ }\bibfield  {title} {\enquote {\bibinfo {title} {{Recovering Galaxy
  Cluster Convergence from Lensed CMB with Generative Adversarial Networks}},}\
  }\href {\doibase 10.48550/arXiv.2211.08990} {\bibfield  {journal} {\bibinfo
  {journal} {arXiv e-prints}\ ,\ \bibinfo {eid} {arXiv:2211.08990}} (\bibinfo
  {year} {2022})},\ \Eprint {http://arxiv.org/abs/2211.08990} {arXiv:2211.08990
  [astro-ph.CO]} \BibitemShut {NoStop}%
\bibitem [{\citenamefont {Foreman}\ \emph {et~al.}(2023)\citenamefont
  {Foreman}, \citenamefont {Hotinli}, \citenamefont {Madhavacheril},
  \citenamefont {van Engelen},\ and\ \citenamefont
  {Kreisch}}]{Foreman:2022ves}%
  \BibitemOpen
  \bibfield  {author} {\bibinfo {author} {\bibfnamefont {Simon}\ \bibnamefont
  {Foreman}}, \bibinfo {author} {\bibfnamefont {Selim~C.}\ \bibnamefont
  {Hotinli}}, \bibinfo {author} {\bibfnamefont {Mathew~S.}\ \bibnamefont
  {Madhavacheril}}, \bibinfo {author} {\bibfnamefont {Alexander}\ \bibnamefont
  {van Engelen}}, \ and\ \bibinfo {author} {\bibfnamefont {Christina~D.}\
  \bibnamefont {Kreisch}},\ }\bibfield  {title} {\enquote {\bibinfo {title}
  {{Subtracting the kinetic Sunyaev-Zeldovich effect from the cosmic microwave
  background with surveys of large-scale structure}},}\ }\href {\doibase
  10.1103/PhysRevD.107.083502} {\bibfield  {journal} {\bibinfo  {journal}
  {Phys. Rev. D}\ }\textbf {\bibinfo {volume} {107}},\ \bibinfo {pages}
  {083502} (\bibinfo {year} {2023})},\ \Eprint
  {http://arxiv.org/abs/2209.03973} {arXiv:2209.03973 [astro-ph.CO]}
  \BibitemShut {NoStop}%
\bibitem [{\citenamefont {{Li}}\ \emph {et~al.}(2022)\citenamefont {{Li}},
  \citenamefont {{Ilayda Onur}}, \citenamefont {{Dodelson}},\ and\
  \citenamefont {{Chaudhari}}}]{2022arXiv220507368L}%
  \BibitemOpen
  \bibfield  {author} {\bibinfo {author} {\bibfnamefont {Peikai}\ \bibnamefont
  {{Li}}}, \bibinfo {author} {\bibfnamefont {Ipek}\ \bibnamefont {{Ilayda
  Onur}}}, \bibinfo {author} {\bibfnamefont {Scott}\ \bibnamefont
  {{Dodelson}}}, \ and\ \bibinfo {author} {\bibfnamefont {Shreyas}\
  \bibnamefont {{Chaudhari}}},\ }\bibfield  {title} {\enquote {\bibinfo {title}
  {{High-Resolution CMB Lensing Reconstruction with Deep Learning}},}\ }\href
  {\doibase 10.48550/arXiv.2205.07368} {\bibfield  {journal} {\bibinfo
  {journal} {arXiv e-prints}\ ,\ \bibinfo {eid} {arXiv:2205.07368}} (\bibinfo
  {year} {2022})},\ \Eprint {http://arxiv.org/abs/2205.07368} {arXiv:2205.07368
  [astro-ph.CO]} \BibitemShut {NoStop}%
\bibitem [{\citenamefont {Han}\ and\ \citenamefont
  {Sehgal}(2022)}]{Han:2021vtm}%
  \BibitemOpen
  \bibfield  {author} {\bibinfo {author} {\bibfnamefont {Dongwon}\ \bibnamefont
  {Han}}\ and\ \bibinfo {author} {\bibfnamefont {Neelima}\ \bibnamefont
  {Sehgal}},\ }\bibfield  {title} {\enquote {\bibinfo {title} {{Mitigating
  foreground bias to the CMB lensing power spectrum for a CMB-HD survey}},}\
  }\href {\doibase 10.1103/PhysRevD.105.083516} {\bibfield  {journal} {\bibinfo
   {journal} {Phys. Rev. D}\ }\textbf {\bibinfo {volume} {105}},\ \bibinfo
  {pages} {083516} (\bibinfo {year} {2022})},\ \Eprint
  {http://arxiv.org/abs/2112.02109} {arXiv:2112.02109 [astro-ph.CO]}
  \BibitemShut {NoStop}%
\bibitem [{\citenamefont {Hotinli}\ \emph {et~al.}(2022)\citenamefont
  {Hotinli}, \citenamefont {Meyers}, \citenamefont {Trendafilova},
  \citenamefont {Green},\ and\ \citenamefont {van Engelen}}]{Hotinli:2021umk}%
  \BibitemOpen
  \bibfield  {author} {\bibinfo {author} {\bibfnamefont {Selim~C.}\
  \bibnamefont {Hotinli}}, \bibinfo {author} {\bibfnamefont {Joel}\
  \bibnamefont {Meyers}}, \bibinfo {author} {\bibfnamefont {Cynthia}\
  \bibnamefont {Trendafilova}}, \bibinfo {author} {\bibfnamefont {Daniel}\
  \bibnamefont {Green}}, \ and\ \bibinfo {author} {\bibfnamefont {Alexander}\
  \bibnamefont {van Engelen}},\ }\bibfield  {title} {\enquote {\bibinfo {title}
  {{The benefits of CMB delensing}},}\ }\href {\doibase
  10.1088/1475-7516/2022/04/020} {\bibfield  {journal} {\bibinfo  {journal}
  {JCAP}\ }\textbf {\bibinfo {volume} {04}},\ \bibinfo {pages} {020} (\bibinfo
  {year} {2022})},\ \Eprint {http://arxiv.org/abs/2111.15036} {arXiv:2111.15036
  [astro-ph.CO]} \BibitemShut {NoStop}%
\bibitem [{\citenamefont {Tegmark}\ \emph {et~al.}(2003)\citenamefont
  {Tegmark}, \citenamefont {de~Oliveira-Costa},\ and\ \citenamefont
  {Hamilton}}]{Tegmark:2003ve}%
  \BibitemOpen
  \bibfield  {author} {\bibinfo {author} {\bibfnamefont {Max}\ \bibnamefont
  {Tegmark}}, \bibinfo {author} {\bibfnamefont {Angelica}\ \bibnamefont
  {de~Oliveira-Costa}}, \ and\ \bibinfo {author} {\bibfnamefont {Andrew}\
  \bibnamefont {Hamilton}},\ }\bibfield  {title} {\enquote {\bibinfo {title}
  {{A high resolution foreground cleaned CMB map from WMAP}},}\ }\href
  {\doibase 10.1103/PhysRevD.68.123523} {\bibfield  {journal} {\bibinfo
  {journal} {Phys. Rev. D}\ }\textbf {\bibinfo {volume} {68}},\ \bibinfo
  {pages} {123523} (\bibinfo {year} {2003})},\ \Eprint
  {http://arxiv.org/abs/astro-ph/0302496} {arXiv:astro-ph/0302496} \BibitemShut
  {NoStop}%
\bibitem [{\citenamefont {Verde}\ \emph {et~al.}(2019)\citenamefont {Verde},
  \citenamefont {Treu},\ and\ \citenamefont {Riess}}]{Verde:2019ivm}%
  \BibitemOpen
  \bibfield  {author} {\bibinfo {author} {\bibfnamefont {L.}~\bibnamefont
  {Verde}}, \bibinfo {author} {\bibfnamefont {T.}~\bibnamefont {Treu}}, \ and\
  \bibinfo {author} {\bibfnamefont {A.~G.}\ \bibnamefont {Riess}},\ }\bibfield
  {title} {\enquote {\bibinfo {title} {{Tensions between the Early and the Late
  Universe}},}\ }\href {\doibase 10.1038/s41550-019-0902-0} {\bibfield
  {journal} {\bibinfo  {journal} {Nature Astron.}\ }\textbf {\bibinfo {volume}
  {3}},\ \bibinfo {pages} {891} (\bibinfo {year} {2019})},\ \Eprint
  {http://arxiv.org/abs/1907.10625} {arXiv:1907.10625 [astro-ph.CO]}
  \BibitemShut {NoStop}%
\bibitem [{\citenamefont {Di~Valentino}\ \emph {et~al.}(2021)\citenamefont
  {Di~Valentino} \emph {et~al.}}]{DiValentino:2020vvd}%
  \BibitemOpen
  \bibfield  {author} {\bibinfo {author} {\bibfnamefont {Eleonora}\
  \bibnamefont {Di~Valentino}} \emph {et~al.},\ }\bibfield  {title} {\enquote
  {\bibinfo {title} {{Cosmology intertwined III: $f\sigma_8$ and $S_8$}},}\
  }\href {\doibase 10.1016/j.astropartphys.2021.102604} {\bibfield  {journal}
  {\bibinfo  {journal} {Astropart. Phys.}\ }\textbf {\bibinfo {volume} {131}},\
  \bibinfo {pages} {102604} (\bibinfo {year} {2021})},\ \Eprint
  {http://arxiv.org/abs/2008.11285} {arXiv:2008.11285 [astro-ph.CO]}
  \BibitemShut {NoStop}%
\bibitem [{\citenamefont {Perivolaropoulos}\ and\ \citenamefont
  {Skara}(2022)}]{Perivolaropoulos:2021jda}%
  \BibitemOpen
  \bibfield  {author} {\bibinfo {author} {\bibfnamefont {Leandros}\
  \bibnamefont {Perivolaropoulos}}\ and\ \bibinfo {author} {\bibfnamefont
  {Foteini}\ \bibnamefont {Skara}},\ }\bibfield  {title} {\enquote {\bibinfo
  {title} {{Challenges for \ensuremath{\Lambda}CDM: An update}},}\ }\href
  {\doibase 10.1016/j.newar.2022.101659} {\bibfield  {journal} {\bibinfo
  {journal} {New Astron. Rev.}\ }\textbf {\bibinfo {volume} {95}},\ \bibinfo
  {pages} {101659} (\bibinfo {year} {2022})},\ \Eprint
  {http://arxiv.org/abs/2105.05208} {arXiv:2105.05208 [astro-ph.CO]}
  \BibitemShut {NoStop}%
\bibitem [{\citenamefont {Peirani}\ \emph {et~al.}(2017)\citenamefont
  {Peirani}, \citenamefont {Dubois}, \citenamefont {Volonteri}, \citenamefont
  {Devriendt}, \citenamefont {Bundy}, \citenamefont {Silk}, \citenamefont
  {Pichon}, \citenamefont {Kaviraj}, \citenamefont {Gavazzi},\ and\
  \citenamefont {Habouzit}}]{Peirani:2016qvp}%
  \BibitemOpen
  \bibfield  {author} {\bibinfo {author} {\bibfnamefont {S.}~\bibnamefont
  {Peirani}}, \bibinfo {author} {\bibfnamefont {Y.}~\bibnamefont {Dubois}},
  \bibinfo {author} {\bibfnamefont {M.}~\bibnamefont {Volonteri}}, \bibinfo
  {author} {\bibfnamefont {J.}~\bibnamefont {Devriendt}}, \bibinfo {author}
  {\bibfnamefont {K.}~\bibnamefont {Bundy}}, \bibinfo {author} {\bibfnamefont
  {J.}~\bibnamefont {Silk}}, \bibinfo {author} {\bibfnamefont {C.}~\bibnamefont
  {Pichon}}, \bibinfo {author} {\bibfnamefont {S.}~\bibnamefont {Kaviraj}},
  \bibinfo {author} {\bibfnamefont {R.}~\bibnamefont {Gavazzi}}, \ and\
  \bibinfo {author} {\bibfnamefont {M.}~\bibnamefont {Habouzit}},\ }\bibfield
  {title} {\enquote {\bibinfo {title} {{Density profile of dark matter haloes
  and galaxies in the horizon\textendash{}agn simulation: the impact of AGN
  feedback}},}\ }\href {\doibase 10.1093/mnras/stx2099} {\bibfield  {journal}
  {\bibinfo  {journal} {Mon. Not. Roy. Astron. Soc.}\ }\textbf {\bibinfo
  {volume} {472}},\ \bibinfo {pages} {2153--2169} (\bibinfo {year} {2017})},\
  \Eprint {http://arxiv.org/abs/1611.09922} {arXiv:1611.09922 [astro-ph.GA]}
  \BibitemShut {NoStop}%
\bibitem [{\citenamefont {Sorini}\ \emph {et~al.}(2022)\citenamefont {Sorini},
  \citenamefont {Dave}, \citenamefont {Cui},\ and\ \citenamefont
  {Appleby}}]{Sorini:2021eac}%
  \BibitemOpen
  \bibfield  {author} {\bibinfo {author} {\bibfnamefont {Daniele}\ \bibnamefont
  {Sorini}}, \bibinfo {author} {\bibfnamefont {Romeel}\ \bibnamefont {Dave}},
  \bibinfo {author} {\bibfnamefont {Weiguang}\ \bibnamefont {Cui}}, \ and\
  \bibinfo {author} {\bibfnamefont {Sarah}\ \bibnamefont {Appleby}},\
  }\bibfield  {title} {\enquote {\bibinfo {title} {{How baryons affect haloes
  and large-scale structure: a unified picture from the Simba simulation}},}\
  }\href {\doibase 10.1093/mnras/stac2214} {\bibfield  {journal} {\bibinfo
  {journal} {Mon. Not. Roy. Astron. Soc.}\ }\textbf {\bibinfo {volume} {516}},\
  \bibinfo {pages} {883--906} (\bibinfo {year} {2022})},\ \Eprint
  {http://arxiv.org/abs/2111.13708} {arXiv:2111.13708 [astro-ph.GA]}
  \BibitemShut {NoStop}%
\bibitem [{\citenamefont {Chung}\ \emph {et~al.}(2020)\citenamefont {Chung},
  \citenamefont {Foreman},\ and\ \citenamefont {van Engelen}}]{Chung:2019bsk}%
  \BibitemOpen
  \bibfield  {author} {\bibinfo {author} {\bibfnamefont {Eegene}\ \bibnamefont
  {Chung}}, \bibinfo {author} {\bibfnamefont {Simon}\ \bibnamefont {Foreman}},
  \ and\ \bibinfo {author} {\bibfnamefont {Alexander}\ \bibnamefont {van
  Engelen}},\ }\bibfield  {title} {\enquote {\bibinfo {title} {{Baryonic
  effects on CMB lensing and neutrino mass constraints}},}\ }\href {\doibase
  10.1103/PhysRevD.101.063534} {\bibfield  {journal} {\bibinfo  {journal}
  {Phys. Rev. D}\ }\textbf {\bibinfo {volume} {101}},\ \bibinfo {pages}
  {063534} (\bibinfo {year} {2020})},\ \bibinfo {note} {[Erratum: Phys.Rev.D
  102, 109903 (2020)]},\ \Eprint {http://arxiv.org/abs/1910.09565}
  {arXiv:1910.09565 [astro-ph.CO]} \BibitemShut {NoStop}%
\bibitem [{\citenamefont {{Hui}}\ \emph {et~al.}(2007)\citenamefont {{Hui}},
  \citenamefont {{Gazta{\~n}aga}},\ and\ \citenamefont
  {{Loverde}}}]{Hui:2007cu}%
  \BibitemOpen
  \bibfield  {author} {\bibinfo {author} {\bibfnamefont {Lam}\ \bibnamefont
  {{Hui}}}, \bibinfo {author} {\bibfnamefont {Enrique}\ \bibnamefont
  {{Gazta{\~n}aga}}}, \ and\ \bibinfo {author} {\bibfnamefont {Marilena}\
  \bibnamefont {{Loverde}}},\ }\bibfield  {title} {\enquote {\bibinfo {title}
  {{Anisotropic magnification distortion of the 3D galaxy correlation. I. Real
  space}},}\ }\href {\doibase 10.1103/PhysRevD.76.103502} {\bibfield  {journal}
  {\bibinfo  {journal} {\prd}\ }\textbf {\bibinfo {volume} {76}},\ \bibinfo
  {eid} {103502} (\bibinfo {year} {2007})},\ \Eprint
  {http://arxiv.org/abs/0706.1071} {arXiv:0706.1071 [astro-ph]} \BibitemShut
  {NoStop}%
\bibitem [{\citenamefont {Green}\ \emph {et~al.}(2017)\citenamefont {Green},
  \citenamefont {Meyers},\ and\ \citenamefont {van Engelen}}]{Green:2016cjr}%
  \BibitemOpen
  \bibfield  {author} {\bibinfo {author} {\bibfnamefont {Daniel}\ \bibnamefont
  {Green}}, \bibinfo {author} {\bibfnamefont {Joel}\ \bibnamefont {Meyers}}, \
  and\ \bibinfo {author} {\bibfnamefont {Alexander}\ \bibnamefont {van
  Engelen}},\ }\bibfield  {title} {\enquote {\bibinfo {title} {{CMB Delensing
  Beyond the B Modes}},}\ }\href {\doibase 10.1088/1475-7516/2017/12/005}
  {\bibfield  {journal} {\bibinfo  {journal} {JCAP}\ }\textbf {\bibinfo
  {volume} {12}},\ \bibinfo {pages} {005} (\bibinfo {year} {2017})},\ \Eprint
  {http://arxiv.org/abs/1609.08143} {arXiv:1609.08143 [astro-ph.CO]}
  \BibitemShut {NoStop}%
\end{thebibliography}%
\end{document}